\documentclass{statsoc}
\usepackage{amsmath,amssymb}
\usepackage[a4paper]{geometry}
\usepackage{enumerate}
\usepackage{natbib}
\usepackage{subfigure}
\usepackage{graphicx}
\usepackage{etoolbox}
\usepackage{color}
\usepackage{multirow}
\usepackage{multicol}
\usepackage{float}
\usepackage{bm}
\usepackage{url}
\newtheorem{theorem}{Theorem}
\newtheorem{definition}{Definition}

\newtheorem{remark}{Remark}
\newtheorem{example}{Example}
\newtheorem{assumption}{Assumption}
\newtheorem{proposition}{Proposition}

 \newtheorem{innercustomthm}{Theorem}
\newenvironment{customthm}[1]
  {\renewcommand\theinnercustomthm{#1}\innercustomthm}
  {\endinnercustomthm}

\newtheorem{innercustomprop}{Proposition}
\newenvironment{customprop}[1]
  {\renewcommand\theinnercustomprop{#1}\innercustomprop}
  {\endinnercustomprop}

\usepackage{mathtools}
\DeclarePairedDelimiter{\ceil}{\lceil}{\rceil}

\title[Semiparametric Model-Based Sensitivity Analysis]{A Semiparametric Approach to Model-Based Sensitivity Analysis in Observational Studies}

\author[B. Zhang and E. J. Tchetgen Tchetgen]{Bo Zhang}
\address{Vaccine and Infectious Disease Division, Fred Hutchinson Cancer Center, Seattle, Washington, U.S.A.}
\email{bzhang3@fredhutch.org}
\author[B. Zhang and E. J. Tchetgen Tchetgen]{Eric J. Tchetgen Tchetgen}
\address{Department of Statistics and Data Science, The Wharton School, University of Pennsylvania, 
Philadelphia, PA, 
U.S.A.}

\begin{document}

\begin{abstract}
When drawing causal inference from observational data, there is almost always concern about unmeasured confounding. One way to tackle this is to conduct a sensitivity analysis. One widely-used sensitivity analysis framework hypothesizes the existence of a scalar unmeasured confounder U and asks how the causal conclusion would change were U measured and included in the primary analysis. Work along this line often makes various parametric assumptions on U, for the sake of mathematical and computational convenience. In this article, we further this line of research by developing a valid sensitivity analysis that leaves the distribution of U unrestricted. Compared to many existing methods in the literature, our method allows for a larger and more flexible family of models, mitigates observable implications (\citealp{franks2019flexible}), and works seamlessly with any primary analysis that models the outcome regression parametrically. We construct both pointwise confidence intervals and confidence bands that are uniformly valid over a given sensitivity parameter space, thus formally accounting for unknown sensitivity parameters. We apply our proposed method on an influential yet controversial study of the causal relationship between war experiences and political activeness using observational data from Uganda.
\end{abstract}


\maketitle

\section{INTRODUCTION}
\label{sec: introduction}
\subsection{Motivating example: War and political participation in Uganda}
\label{subsec: motivating example}
What is the political legacy, if any, of a violent civil war? A tragic observational study in Uganda provides some empirical evidence. In 1988, several failed insurgent groups in northern Uganda were assembled into a new force, called the Lord's Resistance Amy, or LRA. The poverty and unpopularity of the movement lead to its reliance on forced recruitment, or \emph{abduction}. From 1995 to 2004, $60,000$ to $80,000$ youths were estimated to be abducted by LRA for at least a day (\citealp{Annan2006}). About $80\%$ of these abductees escaped, were released, or were rescued after abduction, and many returnees later relocated through a government's ``reception center''(\citealp{Blattman2009}). 

To better understand the effects and consequences of war experiences, a representative survey of male youth in eight rural subcounties in Uganda was conducted during 2005 to 2006. In particular, \cite{Blattman2009} studied the causal link from war experiences to political engagement using evidence from the data, and found that abduction leads to an $11.0$ percentage point increase in the probability that a youth over $18$ years old voted in the $2005$ referendum on restoring multi-party politics. This result is of particular interest as it defies expectations: political scientists often worry that ex-combatants face a lifetime of crime and banditry, and remain alienated and ``at war" in their own minds (\citealp{Blattman2009,spear2016disarmament}), which makes rebuilding of the society much more challenging after conflict, and could contribute to the well-known ``conflict trap” (\citealp{Collier2008}). \citet{Blattman2009}'s empirical study offered some encouraging evidence that war experiences could lead to greater postwar political engagement.

Throughout the analysis, \cite{Blattman2009} assumes ``conditional unconfoundedness,'' i.e., abduction is effectively randomized conditional on the observed covariates. Many sources of bias exist, as acknowledged by the author. For instance, the observed causal relationship could be spurious if more politically active young men were targeted by the LRA, and this ``political activeness'' was not measured and adjusted for. To address this concern, the author conducted a ``thought experiment,'' or a sensitivity analysis, following the framework described in \citet{Rosenbaum1983}  and \citet{Imbens2003}. According to their framework, an independent binary unmeasured confounder $U \sim \text{Bernoulli}(0.5)$ is hypothesized to exist and it is asked how the causal conclusions would change were this $U$ measured and included in the analysis, in addition to the collected observed covariates. Specifically, the following model is considered: 
\begin{equation}
\begin{split}
    &U \mid \mathbf{X} \sim \text{Bernoulli}~(0.5),\\
    &\text{logit}(Y \mid Z, \mathbf{X}, U) = \beta Z + \mathbf\lambda^T \mathbf{X} + c_\delta U, \\
    &\text{logit}(Z \mid \mathbf{X}, U) = \mathbf\kappa^T \mathbf{X} + c_\gamma U,
\end{split}
\label{model: imbens sens model}
\end{equation}
where $U$ is a hypothesized binary unmeasured confounder (e.g., $U = 1$ if the subject is politically active and $0$ otherwise), $\mathbf{X}$ a vector of measured covariates including the intercept, $Z$ the binary treatment (having been abducted), $Y$ the binary response (whether or not the subject voted in the 2005 referendum), and $\beta$ the treatment effect on the logit scale. In Model \eqref{model: imbens sens model}, $(c_\delta, c_\gamma)$ are sensitivity parameters: $c_\gamma$ quantifies the association between the treatment assignment $Z$ and the hypothetical unmeasured confounder $U$, and $c_\delta$ the association between the outcome $Y$ and $U$. For any fixed pair of sensitivity parameters $(c_\delta, c_\gamma)$, the observed data likelihood of Model \eqref{model: imbens sens model} is maximized, and the $100\times(1-\alpha)\%$ confidence interval for $\beta$ is reported. In the above specification, the parametric model for the outcome $Y$ is inherited from the primary analysis assuming ``no unmeasured confounding'' corresponding to $(c_\delta, c_\gamma) = (0, 0)$. As noted by \cite{Imbens2003}, the model specification can be readily modified or extended conceptually. We will refer to the class of sensitivity analysis methods that hypothesize the existence of an unmeasured confounder as the \emph{added-variable approach}, or \emph{omitted-variable approach} (\citealp{Wooldridge2008}), to sensitivity analysis in this paper.

\subsection{Limitations of \citet{Rosenbaum1983} and \citet{Imbens2003}'s models}
Added or omitted-variable approaches to sensitivity analysis, e.g., Model \eqref{model: imbens sens model}, have at least three desirable features. First, they are seamlessly integrated to the primary analysis based on modeling potential outcomes. In fact, the primary analysis is restored by setting $(c_\delta, c_\gamma) = (0, 0)$. Second, sensitivity parameters of the model are intuitive, transparent, and easy to communicate, and the number of sensitivity parameters is small. Third, when empirical researchers have in mind some particular unmeasured confounder $U = U^\ast$ and can specify the distribution in the population, say from some external data source, Model (\ref{model: imbens sens model}) can be directly employed to assess robustness of causal conclusions to such an unmeasured confounder. 

However, having conducted a sensitivity analysis under Model \eqref{model: imbens sens model}, one natural question to ponder on is the following: What role does the parametric assumption on $U$ play in statistical inference? After all, $U$ is not observed and it may be preferable not to impose any parametric assumptions on the distribution of this unobserved component. A related and even more concerning feature to some researchers is that specifying $U \sim \text{Bernoulli}(0.5)$ as in \eqref{model: imbens sens model} introduces too strong \emph{observable implications} (\citealp{franks2019flexible}): The observed data $Y \mid Z, \mathbf{X}$ is distributed as a two-component mixture of logistic regressions with equal weights after integrating out the binary unmeasured confounder $U$. This may not be flexible enough to describe the data at hand, and makes sensitivity parameters easily identified from data, as acknowledged by many authors (\citealp{Copas1997,Scharfstein1999,Imbens2003}). 

In addition to these theoretical and philosophical concerns, a more important question of practical relevance emerges: Is it possible that the parametric assumption on $U$ somehow colludes with the data at hand to produce a more favorable sensitivity analysis result? Even in scenarios where empirical researchers have in mind, or are encouraged by the scientific community to consider the possibility of bias due to a specific unmeasured confounder $U^\ast$, often little ``prior knowledge'' is available to correctly specify the distribution of $U^\ast \mid \mathbf{X}$ in the population. In some circumstances, sensitivity analysis conclusions can be quite sensitive to parametric assumptions regarding the distribution of $U$. For instance, in a study of the effect of second-hand smoking on blood-lead levels, \citet{zhang2018calibrated} proposed that attending a public versus private school could be an important binary unmeasured confounder $U$ in their analysis and they found that the causal conclusion could be explained away when $U \sim \text{Bernoulli}(0.5)$ and $(c_\delta, c_\gamma) = (1.2, 1.2)$, but not when $U \sim \text{Bernoulli}(0.1)$ and $(c_\delta, c_\gamma)$ as large as $(2.0, 2.0)$. Consider two independent study units $i$ and $j$ with the same observed covariates but a possibly different unmeasured confounder $U$ as in \citet{Cornfield1959} and \citet{Rosenbaum2002a}. Their odds ratio of being exposed to second-hand smoking is $\text{OR} = \exp\{c_\gamma(u_i - u_j)\}$, which has an expected value of $1.91$ when $U \sim \text{Bernoulli}(0.5)$ and $c_\gamma = 1.2$ but only $1.68$ when $U \sim \text{Bernoulli}(0.1)$ and $c_\gamma = 2.0$. It is unclear which if any of these results one should believe.

\subsection{A semiparametric model-based sensitivity analysis}
Legitimate concerns regarding Model \eqref{model: imbens sens model} motivate us to develop a method that still preserves key elements that have made \citet{Rosenbaum1983} and \citet{Imbens2003}'s original proposals popular, while avoiding unfounded and untestable parametric assumptions typically made about the distribution of $U$, therefore allowing for the latter to remain unrestricted and mitigating undesirable observable implications (\citealp{franks2019flexible}) of such unnecessary restrictions. 

We leverage modern semiparametric theory (\citealp{Newey1990,Bickel1993efficient,van2000asymptotic,Tsiatis2006}) to construct a consistent and asymptotically normal (CAN) estimator of the average treatment effect in a model where the outcome regression model and the propensity score model are correctly specified, while the distribution of the hypothesized unmeasured confounder is unrestricted. We leverage this result to develop a two-parameter sensitivity analysis. An important feature of the proposed estimator is that it attains the efficiency bound for the semiparametric model whenever a working model for distribution of $U$ is correct, yet it is robust to possible misspecification of such a model as it remains consistent and asymptotically normal in such an eventuality.



Our proposal aims to strike a balance between generality and complexity (\citealp{VanderWeele2011}). The proposed approach is general in the following sense. First, it does not place distributional assumptions on the unmeasured confounder $U$. Second, it works seamlessly with any parametric outcome regression model $\mathbb{E}[Y \mid Z, \mathbf{X}]$ that empirical researchers routinely fit in their primary analysis. For instance, in the example of \cite{Blattman2009}, a probit or logistic regression relating the binary outcome, the binary treatment, and baseline covariates is fit in the primary analysis. Our proposed method would directly build upon this model specification in the primary analysis by inserting a hypothesized unmeasured confounder $U$ with unrestricted distribution in the population. More importantly, our proposed approach remains practical and easy-to-use and does not sacrifice the lucidity and transparency of the original widely used method proposed by \cite{Rosenbaum1983} and generalized by \cite{Imbens2003}. 

The rest of the paper is organized as follows. In Section \ref{sec: notaion assumption}, we review key notation, assumptions, and background on sensitivity analysis in observational studies, with emphasis on the added or omitted-variable approach. In Section \ref{sec: semi model} and \ref{sec: estimation and inference}, we introduce key concepts of semiparametric theory, specify our semiparametric model, and describe estimation and inference procedures. We present extensive simulation results in Section \ref{sec: first set of simulation}. We discuss how to report a sensitivity analysis in Section \ref{sec: one-param SA} and how to interpret the result in Section \ref{sec: interpret}. The proposed method is applied to the war and political participation study in Section \ref{sec: application} and Section \ref{sec: conclude} concludes with a brief discussion. Relevant data and \textsf{R} code to reproduce results in this paper is available at \url{https://github.com/bzhangupenn/Code_for_reproducing_semi_SA}. 



\section{NOTATION AND LITERATURE REVIEW}
\label{sec: notaion assumption}
\subsection{Notation and assumption}
We briefly review notation and assumptions for drawing causal inference from observational studies. Let $Y(z), ~z = 0,1$ be the potential outcome under treatment $Z = z$ (\citealp{Neyman1923}; \citealp{Rubin1974}). This notation implicitly makes the \emph{stable unit treatment value assumption} (SUTVA) (\citealp{Rubin1980}), i.e., a subject's potential outcome does not depend on the treatment given to others and there is a unique version of treatment defining the intervention of scientific interest. For each subject, we observe data $(\mathbf{X}, Z, Y)$, where $\mathbf{X}$ is a vector of \emph{observed covariates}, $Z$ the \emph{treatment assignment}, and $Y$ the \emph{observed outcome} satisfying $Y = ZY(1) + (1 - Z)Y(0)$. The difference between two mean potential outcomes $\mathbb{E}[Y(1) - Y(0)]$ is called the average treatment effect.

A key assumption in drawing causal inference is the so-called \emph{treatment ignorability assumption} (\citealp{rosenbaum1983central}), also known as the \emph{no unmeasured confounding assumption} (\citealp{Robins1992}), \emph{exchangeability} (\citealp{Greenland1986}), \emph{selection on observables} (\citealp{Barnow1980}), or \emph{treatment exogeneity} (\citealp{Imbens2004}). A version of this assumption states that
\begin{equation*}
     F(y(0), y(1) \mid Z = z, \mathbf{X} = \mathbf{x}) = F(y(0), y(1) \mid \mathbf{X} = \mathbf{x}),~\forall(z, \mathbf{x}),
\end{equation*}
where $F(\cdot)$ denotes the cumulative distribution function. In words, the assumption states that the potential outcomes are jointly independent of the treatment assignment conditional on observed covariates. We further assume that \emph{positivity} holds, i.e., $0 < P(Z = 1 \mid \mathbf{X} = \mathbf{x}) < 1, ~\forall \mathbf{x}$. Under treatment ignorability, some widely used methods for drawing causal inference include: matching (\citealp{Rubin1979}; \citealp{Rosenbaum2002a}; \citealp{Stuart2010}), modeling potential outcomes $\mathbb{E}[Y \mid Z, \mathbf{X}]$ (\citealp{Robins1986}; \citealp{Wasserman1999}; \citealp{Robins2000c}; \citealp{Hill2011}), propensity score weighting and subclassification (\citealp{Rosenbaum1984}; \citealp{Rosenbaum1987b}), g-estimation of a structural nested model (\citealp{Robins1986}; \citealp{Vansteelandt2014}), and doubly robust methods (\citealp{Robins1994}; \citealp{Robins2000b}; \citealp{Bang2005}).

\subsection{Added-variable approach to sensitivity analysis}
\label{subsec: review added variable models}
In many practical scenarios, the ``no unmeasured confounding'' assumption may be a heroic assumption and a major obstacle to drawing valid causal conclusions. Sensitivity analysis is one way to tackle concerns about the potential bias from unmeasured confounding. A sensitivity analysis asks to what extent the causal conclusion drawn from the data at hand would change when the no unmeasured confounding assumption is relaxed. Many sensitivity analysis methods have been proposed for different causal inference frameworks over the years; see, e.g., \citet{Cornfield1959}, \cite{Gastwirth1998}, \cite{Scharfstein1999}, \cite{McCandless2007}, \cite{Ichino2008}, \cite{Rosenbaum1987, Rosenbaum2002a, Rosenbaum2010}, \cite{ding2016sensitivity}, \citet{franks2019flexible}, \cite{zhao_sens_ipw2019}, and \cite{Cinelli2020}, among many others.

One approach to representing unmeasured confounding is to hypothesize the existence of a latent scalar variable $U$ that summarizes unmeasured confounding. The idea is that were $U$ observed and accounted for, there would remain no further unmeasured confounding so that the no unmeasured confounding assumption holds provided one conditions on both $\mathbf{X}$ and $U$ but not otherwise. In order to identify the treatment effect in the presence of this hypothesized unmeasured confounder, the entire data generating process including the distribution of $U$, or at least some aspects of it, is specified. \cite{Rosenbaum1983} first considered the setting of a binary outcome and assumed a discrete stratification variable $S$ such that the treatment assignment is strongly ignorable conditional on $S$ and $U$. \cite{Imbens2003} extended this approach by allowing for continuous measured covariates and considering a normal outcome. The sensitivity analysis model considered in \citet{altonji2005selection} can also be formulated as a version of Model \eqref{model: imbens sens model}. \cite{carnegie2016assessing} further extended the model to a continuous treatment and a normally distributed unmeasured confounder $U$. \citet{Dorie2016} proposed to more flexibly model the response surface using Bayesian Additive Regression Trees (BART), while still assuming that $U$ is an independent binary variable and keeping the parametric specification of the treatment assignment model. More recently, \citet{Cinelli2020} applied the omitted variable bias (OVB) techniques (\citealp{Wooldridge2008}) to constructing a sensitivity analysis for linear structural equation models without specifying the distribution of $U$.


\citet{ding2016sensitivity} developed a two-parameter sensitivity analysis approach called $E$-value. For a binary outcome and a binary treatment, \cite{ding2016sensitivity} showed the true relative risk ratio, even in the presence of unmeasured confounders, is always at least as large as $RR_{ZY\mid\mathbf{x}}^{obs} \big/\frac{RR_{ZU\mid\mathbf{x}} \times RR_{UY\mid\mathbf{x}}}{RR_{ZU\mid\mathbf{x}} + RR_{UY\mid\mathbf{x}} - 1}$, where $RR_{ZY\mid\mathbf{x}}^{obs}$ is the observed risk ratio within stratum $\mathbf{X} = \mathbf{x}$, $RR_{ZU \mid \mathbf{x}}$ the maximal relative risk of $Z$ on $U$ within stratum $\mathbf{X} = \mathbf{x}$, and $RR_{UY \mid \mathbf{x}}$ the maximal relative risk of $U$ on $Y$ within stratum $\mathbf{X} = \mathbf{x}$, with and without treatment. The main appeal of the approach is that it is easy to compute. However, an important limitation of the result is that the correction formally works only on risk ratio scale. Although the authors have developed several approximations to allow for other scales (e.g. odds ratio or additive effects), no formal theoretical guarantees exist as to their inferential correctness. Furthermore, specification of $RR_{ZU \mid \mathbf{x}}$ formally restricts the retrospective likelihood ratio $f(U \mid  Z = 1, \mathbf{X} = \mathbf{x})/f(U \mid Z = 0, \mathbf{X} = \mathbf{x})$ and therefore restricts the retrospective density $f(U \mid Z, \mathbf{X})$. There are two issues with imposing such a restriction; the first issue is that whereas an investigator might have some insight based on background knowledge as to the magnitude of the dependence of $P(Z\mid \mathbf{X}, U)$ on $U$ as it pertains to treatment selection by unobservables (\citealp{Rosenbaum1987}), as we have argued in the introduction,  rarely would she have the level of knowledge about density of $f(U\mid\mathbf{X})$ in order to specify $RR_{ZU \mid \mathbf{x}}$ in a meaningful and easily interpretable manner. Secondly, $RR_{ZU \mid \mathbf{x}}$ does not necessarily accurately encode strength of unmeasured confounding as it can be made arbitrarily large or small (within a certain range) by varying specification of $f(U\mid \mathbf{X})$ while holding $f(Z\mid \mathbf{X}, U)$ fixed. To illustrate, consider the simple case where $U$ is binary and there is no observed covariates $\mathbf{X}$. Fix $Z = \text{expit}(\alpha_0 + \alpha_1 U)$ and it can be shown with straightforward algebra that $RR_{ZU}$ can be made arbitrarily large or small between $\exp(-\alpha_0)$ and $\exp(-\alpha_0 - \alpha_1)$ by varying the ratio $P(U = 1)/P(U = 0)$, a quantity often of limited interest. The approach developed in this paper addresses both limitations of the $E$-value approach.

\section{MODEL SPECIFICATION}
\label{sec: semi model}

\subsection{A semiparametric perspective of the added-variable approach}
\label{subsec: semiparametric model M}
Semiparametric models refer to statistical models where the functional forms of some components of the model are unknown (\citealp{Newey1990,Bickel1993efficient}). As we discuss extensively in the introduction, a natural component to be left unspecified in our setting is the law of the unmeasured confounder. Below, we describe a concrete set-up to be studied in this article.

Consider the full data $D = (\mathbf{X}, U, Z, Y) \overset{i.i.d.}{\sim} \mathcal{P}_D$, where $\mathbf{X}$ is a vector of observed covariates, $U$ a scalar unmeasured confounder, $Z$ the treatment and $Y$ the response. The observed data $O$ only consist of $(\mathbf{X}, Z, Y)$ as $U$ is not observed. We factor the full data law $\mathcal{P}_D$ as follows:
\[
f(Y, Z, \mathbf{X}, U) = f(Y \mid Z, \mathbf{X}, U) \cdot f(Z \mid \mathbf{X}, U)\cdot f(U \mid \mathbf{X}) \cdot f(\mathbf{X}),
\] 
and consider the following two assumptions on the outcome model $f(Y \mid Z, \mathbf{X}, U)$ and the propensity score model $f(Z \mid \mathbf{X}, U)$:


\begin{assumption}\rm
The outcome model relating $Y$ to $Z$, $\mathbf{X}$, and $U$ satisfies $\mathbb{E}[Y\mid Z, \mathbf{X}, U] = g_1^{-1}( \beta Z + \lambda^T \mathbf{X} + c_\delta U)$, and $Y$ belongs to exponential family with canonical link function $g_1$.
\label{assump: outcome model correct}
\end{assumption}

\begin{assumption}\rm
The propensity score model relating $Z$ to $\mathbf{X}$ and $U$ satisfies $\mathbb{E}[Z \mid \mathbf{X}, U] = g_2^{-1}(\kappa^T \mathbf{X} + c_\gamma U)$, and $Z$ belongs to exponential family with canonical link function $g_2$.
\label{assump: pscore model correct}
\end{assumption}

Assumption \ref{assump: outcome model correct} states that the effect of $U$ on $Y$ is additive on the scale defined by the link function $g_1$ and excludes any $ZU$ interaction. As discussed in Section \ref{sec: introduction}, the outcome model specification inherits that in a primary analysis assuming no unmeasured confounding. Similarly, Assumption \ref{assump: pscore model correct} states that the effect of $U$ on $Z$ is additive on the scale defined by the link function $g_2$. 

\begin{remark}\rm
The method developed in this article can be immediately extended to models with $ZU$ interaction by incorporating an additional sensitivity parameter characterizing $ZU$'s effect on $Y$. We focus on the model where $U$ does not interact with $Z$ because it involves fewer sensitivity parameters, is easier to interpret, and is widely adopted in the literature; see, e.g., \citet{Rosenbaum1983}, \citet{Imbens2003}, and \citet{Rosenbaum2002a}.
\end{remark}

\begin{remark}\rm
It will be clear later when we construct the semiparametric estimator that it is not strictly required to posit exponential family models. We focus on this family of models because they are familiar to empirical researchers and routinely used in practice.
\end{remark}

To summarize, we consider making inference about the $q$-dimensional parameters $\theta = (\lambda, \beta, \kappa)$ in the following semiparametric model $\mathcal{M}_{c_\delta, c_\gamma}$ indexed by the fixed sensitivity parameters $(c_\delta, c_\gamma)$:
\begin{equation}
    \begin{split}
    \mathcal{M}_{c_\delta, c_\gamma} ~:= ~&(U, \mathbf{X}) \sim F(\cdot), ~ F(\cdot) ~\text{is an unrestricted law},\\
    &P(Y = y \mid Z = z, \mathbf{X} = \mathbf{x}, U = u) = f(y \mid z, \mathbf{x}, u; \lambda, \beta, c_\delta), \\
    &P(Z = z \mid \mathbf{X} = \mathbf{x}, U = u) = f(z \mid \mathbf{x}, u; \kappa, c_\gamma).
    \end{split}
\end{equation}
In words, $\mathcal{M}_{c_\delta, c_\gamma}$ represents a semiparametric model where both the outcome model and the propensity score model are correctly specified, with \emph{known} association between $U$ and $Y$ and between $U$ and $Z$, and unrestricted joint law of $(U, \mathbf{X})$. For notational simplicity, we suppress the dependence on $(c_\delta, c_\gamma)$ in the rest of the article and write $\mathcal{M}$ in place of $\mathcal{M}_{c_\delta, c_\gamma}$. Semiparametric model $\mathcal{M}$ contains widely used models proposed by \citet{Rosenbaum1983} and \citet{Imbens2003}. 



\subsection{Identification of sensitivity parameters}
We discuss identification results in this section. For simplicity, we only consider the situation where observed covariates $\mathbf{X}$ are omitted, and $(Y, Z, U)$ are all binary. Consider the following saturated models for $Z$ and $Y$:
\begin{equation*}
\begin{split}
    &f(Y = 1 \mid Z, U) = \text{expit}\{\beta_0 + \beta_z Z + \beta_u U + \beta_{zu} ZU\}, \\
    &f(Z = 1 \mid U) = \text{expit}\{\alpha_0 + \alpha_u U\}.
    \end{split}
\end{equation*}
Let us further parametrize $U$ by $f(U; \xi)$ so that the probability of jointly observing $Y = y$ and $Z = z$ is given by 
\begin{equation*}
\begin{split}
    f(Y = y, Z = z) := &f(Y = y, Z = z; \beta_0, \beta_z, \beta_u, \beta_{zu}, \alpha_0, \alpha_u, \xi) \\
    = &\int f(y \mid z, u; \beta_0, \beta_z, \beta_u, \beta_{zu})\cdot f(z \mid u; \alpha_0, \alpha_u)\cdot f(u; \xi)~du.
\end{split}
\end{equation*}

Since both $Y$ and $Z$ are binary, there are only three degrees of freedom, namely $f(Y = 1, Z = 1)$, $f(Y = 1, Z = 0)$, and $f(Y = 0, Z = 1)$. There are more unknown parameters than degrees of freedom so the model cannot be identified without further restrictions. Assumption \ref{assump: outcome model correct} says $U$'s effect on $Y$ is linear and equal to $c_\delta$, which implies $\beta_{zu} = 0$ and $\beta_u = c_\delta$. Similarly, Assumption \ref{assump: pscore model correct} says $\alpha_u = c_\gamma$. Under Assumption \ref{assump: outcome model correct} and \ref{assump: pscore model correct}, probability of jointly observing $Y = y$ and $Z = z$ then reduces to
\begin{equation*}
    f(Y = y, Z = z; \beta_0, \beta_z, \alpha_0, \xi) = \int \text{expit}\{\beta_0 + \beta_z z + c_\delta u\}\cdot\text{expit}\{\alpha_0 + c_\gamma u\}\cdot f(u; \xi)~du.
\end{equation*}

Proposition \ref{prop: identification} states an identification result in this case.

\begin{proposition}\rm
\label{prop: identification}
Let $Y$, $Z$, and $U$ be binary. Suppose that there is no $Z, U$ interaction in the outcome model and that $c_\delta$ and $c_\gamma$ are fixed sensitivity parameters. Then it is true that
\[
f(Y = y, Z = z; \beta^{(1)}_0, \beta^{(1)}_z, \alpha^{(1)}_0, \xi) = f(Y = y, Z = z; \beta^{(2)}_0, \beta^{(2)}_z, \alpha^{(2)}_0, \xi)
\] implies $\beta^{(1)}_0 = \beta^{(2)}_0$, $\beta^{(1)}_z = \beta^{(2)}_z$, and $\alpha^{(1)}_0 = \alpha^{(2)}_0$, for all $\xi$.
\label{thm: identification when binary}
\end{proposition}

All proofs in this article are left to Supplementary Material C.

Proposition \ref{prop: identification} essentially says that for \emph{fixed} sensitivity parameters $(c_\delta, c_\gamma)$ and any distribution of $U$ parametrized by $\xi$, parameters $(\beta_0, \beta_a, \alpha_0)$ could be uniquely identified in the simple case with no observed covariates and binary $(Y, Z, U)$. If $(c_\delta, c_\gamma)$ are left unspecified, then $(\beta_0, \beta_a, \alpha_0, c_\gamma, c_\delta)$ cannot be jointly identified from the observed data with only three degrees of freedom. In other words, $(c_\delta, c_\gamma)$ should indeed be treated as sensitivity parameters rather than structural parameters to be identified in this simple case.

In more general cases, with parametric assumptions to incorporate observed covariates $\mathbf{X}$, sensitivity parameters $(c_\delta, c_\gamma)$ may become identifiable from the observed data (\citealp{Copas1997, franks2019flexible}). However, the identification is much weaker compared to positing parametric assumptions of the distribution of $U$. For instance, under our proposed semiparametric model $\mathcal{M}$ with a normal outcome regression model, the observed law $f(Y \mid Z, \mathbf{X})$ is distributed as a convolution of a normal density and an unknown distribution, instead of a two-component normal mixture as in Model \eqref{model: imbens sens model}.

\section{ESTIMATION AND INFERENCE}
\label{sec: estimation and inference}
\subsection{Influence functions and estimating equations}
 Most semiparametric theory restricts attention to regular and asymptotically linear (RAL) estimators. An estimator $\hat{\beta}$ for a finite dimensional functional $\beta$ on a statistical model $\mathcal{M}$ (parametric, semiparametric, or nonparametric model) based on i.i.d. data $\{D_i, i = 1,2,...,n \}$ is asymptotically linear if it satisfies 
\begin{equation}
\label{eqn: asymp linear}
    \sqrt{n}(\hat{\beta} - \beta) = \frac{1}{\sqrt{n}}\sum_{i = 1}^n \phi(D_i; \beta) + o_p(1),
\end{equation}
where $\phi(\cdot)$ is often referred to as the influence function of $\hat{\beta}$ and satisfies $\mathbb{E}[\phi(D; \beta)] = 0$ and $\mathbb{E}[\phi^T(D; \beta)\phi(D; \beta)] < \infty$. Regularity is a technical condition that rules out certain ``pathological'' estimators (\citealp{Newey1990}). A regular and asymptotically linear estimator is consistent and asymptotically normal (CAN) with asymptotic covariance matrix $\mathbb{E}[\phi(D; \beta)\phi(D; \beta)^T]$. Within the set of influence functions, there exists an efficient influence function $\phi_{\text{eff}}(D; \beta)$ whose asymptotic variance is no larger than any other influence functions. The variance of $\phi_{\text{eff}}(D; \beta)$ is known as the semiparametric efficiency bound.

Equation \eqref{eqn: asymp linear} suggests a relationship between influence functions and RAL estimators. One general strategy of constructing a semiparametric estimator is to first identify a set containing all influence functions for the semiparametric model, in which case, a candidate RAL estimator can then be obtained by solving the following estimating equation:
\[
\mathbb{P}_n\{\hat{\text{IF}}(D, \hat{\beta})\} = 0,
\] where $\hat{\text{IF}}$ is an estimate of the influence function obtained under $\mathcal{M}$. Under certain regularity conditions, it will then typically be the case that $\hat{\beta}$ thus constructed admits the expansion (\ref{eqn: asymp linear}) with $\phi(\cdot)$ equal to $\text{IF}$. The efficient IF can be obtained by projecting any IF onto the so-called tangent space, defined as the closed linear span of scores for all regular parametric submodels of $\mathcal{M}$ (\citealp{Newey1990,Bickel1993efficient,Robins1994,van2000asymptotic}).


\subsection{Motivating an estimating equation}
We describe how to construct an estimating equation that solves the estimation and associated inference problem of semiparametric model $\mathcal{M}$ described in Section \ref{subsec: semiparametric model M}. Let $\theta = (\lambda, \beta, \kappa)$ denote the finite dimensional parameter of interest in model $\mathcal{M}$,
\[
\mathcal{P}_D = f(Y\mid Z, \mathbf{X}, U)\cdot f(Z \mid \mathbf{X}, U) \cdot f(U\mid \mathbf{X}) \cdot f(\mathbf{X})
\]
the underlying law, and $\mathbb{E}[\cdot]$ expectation taken with respect to $\mathcal{P}_D$. The key obstacle to estimating $\theta$ using the standard likelihood-based methods that maximize the observed data likelihood (e.g., the expectation-maximization (EM) algorithm) is the unspecified component $f(U \mid \mathbf{X})$. To this end, we let $f^\ast(U \mid \mathbf{X}; \xi)$ be a possibly incorrect \emph{working model} for the unknown conditional distribution $f(U \mid \mathbf{X})$, and denote by $\mathbb{E}_\ast[\cdot]$ expectation taken with respect to the joint law 
\[\mathcal{P}_D^\ast = f(Y\mid Z, \mathbf{X}, U)\cdot f(Z \mid \mathbf{X}, U) \cdot f^\ast(U\mid \mathbf{X}; \xi) \cdot f(\mathbf{X}).
\]
Under the joint law $\mathcal{P}_D^\ast$, we may then define the full data score function $S^\ast_\theta(\mathbf{X}, U, Z, Y)$ as the gradient of the log-likelihood of the partially unobserved full data with respect to $\theta$, and calculate the following observed data score function
\begin{equation}
\label{eqn: observed data score}
\begin{split}
    S^\ast_\theta(\mathbf{X}, Z, Y) &= \mathbb{E}_\ast[S^\ast_\theta(\mathbf{X}, U, Z, Y) \mid \mathbf{X}, Z, Y] \\
    &= \frac{\int S^\ast_\theta(Y, Z, \mathbf{X}, u) f(Y, Z \mid \mathbf{X}, u; \theta, c_\delta, c_\gamma)f^\ast(u \mid \mathbf{X}; \xi)~d\mu(u)}{\int f(Y, Z \mid \mathbf{X}, u;\theta, c_\delta, c_\gamma) f^\ast(u \mid \mathbf{X}; \xi) ~d\mu(u)}.
\end{split}
\end{equation}
Unlike the full data score $S^\ast_\theta(\mathbf{X}, U, Z, Y)$ which depends on the unmeasured confounder $U$ and cannot be calculated based on the observed data, the score $S^\ast_\theta(\mathbf{X}, Z, Y)$ depends only on the observed data $(\mathbf{X}, Z, Y)$ and can be readily evaluated under the law $\mathcal{P}_D^\ast$. 

The efficient score is the variation in the score for $\theta$ that is orthogonal to all possible scores of $U \mid \mathbf{X}$, an infinite dimensional space which we characterize in the Supplementary Material A. Specifically, we derive the following observed data efficient score:
\begin{equation}
\label{eqn: efficient observed data score}
    S_{\text{eff}}^\ast(\mathbf{X},Z,Y)  = S^\ast_\theta(\mathbf{X}, Z, Y) - \mathbb{E}_\ast[a(U, \mathbf{X}) \mid \mathbf{X}, Z, Y],
\end{equation}
where $a(U, \mathbf{X})$ satisfies the following constraint:
\begin{equation}
\label{eqn: integral equation}
    \mathbb{E}_\ast[S^\ast_\theta(\mathbf{X}, Z, Y) \mid \mathbf{X}, U] = \mathbb{E}_\ast\{\mathbb{E}_\ast[a(U, \mathbf{X}) \mid \mathbf{X}, Z, Y] \mid \mathbf{X}, U\}.
\end{equation}

One remarkable feature of the efficient score $S_{\text{eff}}^\ast(\mathbf{X},Z,Y)$ is that, although it is calculated under the law $\mathcal{P}_D^\ast$ with a possibly misspecified $f(U\mid\mathbf{X})$ component, it is mean zero under the \emph{true} joint law $\mathcal{P}_D$ by virtue of being orthogonal to any conceivable score for $U \mid \mathbf{X}$ as formalized in the proposition below.

\begin{proposition}\rm
The observed data efficient score $S_{\text{eff}}^\ast(\mathbf{X},Z,Y)$ constructed as in \eqref{eqn: efficient observed data score} and \eqref{eqn: integral equation} satisfies $\mathbb{E}[S_{\text{eff}}^\ast(\mathbf{X},Z,Y) \mid \mathbf{X}, U] = 0$, which implies:
\begin{equation*}
     \mathbb{E}[S_{\text{eff}}^\ast(\mathbf{X},Z,Y)] = 0,
\end{equation*}
where $\mathbb{E}[\cdot]$ is expectation taken with respect to the law $\mathcal{P}_D$.
\label{prop: relate two Hilbert spaces}
\end{proposition}
Proposition \ref{prop: relate two Hilbert spaces} will serve as the basis for constructing the estimating equation. A similar form of robustness of the efficient score to partial misspecification of the nuisance parameter indexing the law of a latent variable has previously appeared  in the context of measurement error (\citealp{Tsiatis2004}), mixed models (\citealp{garcia2016optimal}), and statistical genetics (\citealp{allen2005locally}); however, none of these prior works directly address unmeasured confounding, and therefore to the best of our knowledge the relevance of this type of robustness result is entirely new in the context of sensitivity analysis for unmeasured confounding bias. 

\subsection{Consistency, asymptotic normality, and computation}
\label{subsec: consistency, AN, computation}
Proposition \ref{prop: relate two Hilbert spaces} motivates constructing an estimator for $\theta$ with attractive robustness and efficiency properties by replacing $\mathbb{E}[\cdot]$ with its empirical analogue and forming the following estimating equation:
\begin{equation}
    \label{eqn: estimating equation}
    \sum_{i = 1}^n S_{\text{eff}}^\ast(\mathbf{X}_i, Z_i, Y_i; \theta) = 0.
\end{equation}
Under standard regularity conditions, including nonsingularity of $\mathbb{E}[dS^\ast_{\text{eff}}/d \theta]$ at $\theta$, $\hat{\theta}$ can be shown to be consistent and asymptotically normal as stated in Theorem \ref{thm: CAN}.

\begin{theorem}
\label{thm: CAN}
\rm
Under suitable regularity conditions, the solution $\theta = \hat{\theta}$ to the estimating equation \eqref{eqn: estimating equation} is consistent and asymptotically normal in $\mathcal{M}$, with variance-covariance matrix given by
\[
\mathbf{V} =  (1/n)\cdot\mathbb{E}\{\partial S^\ast_{\text{eff}}(D_i; \theta_0)/\partial \theta\}^{-1} \mathbb{E}\{S^\ast_{\text{eff}}(D_i; \theta_0)S^\ast_{\text{eff}}(D_i; \theta_0)^T\}\mathbb{E}\{\partial S^\ast_{\text{eff}}(D_i; \theta_0)/\partial \theta^T\}^{-1},
\]
where $D_i = (\mathbf{X}_i, Z_i, Y_i)$. If the conditional distribution $f(U \mid \mathbf{X})$ is correctly specified, i.e., when $f^\ast(U \mid \mathbf{X}; \xi) \equiv f(U \mid \mathbf{X})$, then $\hat{\theta}$ is locally efficient with asymptotic variance $\mathbf{V}_\text{eff} = \mathbb{E}[S_{\text{eff}}\cdot S^T_{\text{eff}}]$.
\end{theorem}

To solve the estimating equation \eqref{eqn: estimating equation} using some commonly-used, iterative root-finding algorithm (e.g., the Newton-Raphson method), we need to evaluate the observed data efficient score $S_{\text{eff}}^\ast(\mathbf{X}_i, Z_i, Y_i; \theta^{(k)})$ for each data point $D_i = (\mathbf{X}_i, Z_i, Y_i)$ at the value $\theta^{(k)}$ of the $k$-th iteration. This involves two tasks: (i) solving for $a(U, \mathbf{X})$ at each observed value $\mathbf{X} = \mathbf{X}_i$ so that equation \eqref{eqn: integral equation} holds, and (ii) calculating the observed data efficient score according to \eqref{eqn: efficient observed data score}. In the Supplementary Material B, we describe in detail how to tackle both tasks and give detailed expressions for all quantities involved in the calculation. 

\section{SIMULATION STUDY}
\label{sec: first set of simulation}
In this section, we evaluate performance of our proposed estimator in practice. In particular, we assess two potential sources of bias and error: (i) finite-sample bias as sample size $n \ll \infty$; (ii) approximation errors introduced as the integral equation \eqref{eqn: integral equation} is solved numerically via Tikhonov regularization (regularization parameter $\alpha \gg 0$) and discretization (mesh size parameter $h \gg 0$); see Supplementary Material B for details. This section is planned as follows. In Section \ref{subsec: simulation binary U}, we consider a binary unmeasured confounder $U$, in which case the integral equation \eqref{eqn: integral equation} admits a closed form solution, and we need not be concerned about the approximation error and can focus on assessing the finite-sample bias of the proposed estimator. In Section \ref{subsec: simulation cont U}, we consider a continuous unmeasured confounder $U$ in a setting where $U$ is independent of $\mathbf{X}$ and assess approximation errors. Lastly, Section \ref{subsec: dependent cont U} considers a setting where $U$ is allowed to depend on $\mathbf{X}$. We discuss the computational cost of the proposed algorithm near the end of Section \ref{subsec: simulation cont U}.

\subsection{Binary unmeasured confounder}
\label{subsec: simulation binary U}
We consider a binary $U$ and a binary $Y$ in this section. We compared the proposed semiparametric estimator of $\beta$ to the maximum likelihood estimator obtained via the EM algorithm that treats $U$ as a binary missing covariate with parameter $p$ assumed to equal a specified value; see \citet{zhang2018calibrated} for an implementation of the EM algorithm in this setting. We generated the full data according to the following data-generating process:
\begin{equation}
    \begin{split}
        &X_1 \sim \text{Uniform}(0, 1), ~ X_2 \sim \text{Uniform}(0, 1), \\
        &U \sim \text{Bernoulli}(0.2), \\
        &\text{logit}(Z = 1\mid X_1, X_2, U) = 3X_1 - 3X_2 + c_\gamma U, \\
        &\text{logit}(Y = 1 \mid Z, X_1, X_2, U) = 4X_1 - 4X_2 +  2 Z + c_\delta U,
    \end{split}
    \label{model: simulation model binary U}
\end{equation}
with $c_\delta = c_\gamma = 4$ and sample size $n = 300$, $500$, and $1000$. When $Z$ and $Y$ are both binary, solution to equation \eqref{eqn: integral equation} admits a closed form representation. We used the \textsf{multiroot} function in the \textsf{R} package \textsf{rootSolve} (\citealp{R_rootSolve}) to solve the system of estimating equations. We considered the following four estimators:
\begin{enumerate}
    \item $\hat{\beta}^\ast_{\text{EM}}$: the maximum likelihood estimator with \emph{incorrectly} specified $U \sim \text{Bernoulli}(0.5)$;
    \item $\hat{\beta}_{\text{EM}}$: the maximum likelihood estimator with \emph{correctly} specified $U \sim \text{Bernoulli}(0.2)$;
    \item $\hat{\beta}^\ast_{\text{semi}}$: the semiparametric estimator with \emph{incorrectly} specified $U \sim \text{Bernoulli}(0.5)$;
    \item $\hat{\beta}_{\text{semi}}$: the semiparametric estimator with \emph{correctly} specified $U \sim \text{Bernoulli}(0.2)$.
\end{enumerate}

\begin{table}
\caption{Monte Carlo results of the $4$ estimators: $\hat{\beta}^\ast_{\text{EM}}$, $\hat{\beta}_{\text{EM}}$, $\hat{\beta}^\ast_{\text{semi}}$, $\hat{\beta}_{\text{semi}}$ for various sample sizes: mean, standard error, bias, percentage of bias, coverage, and RMSE. True $\beta$ equals $2.0$.}
\label{tb: binary U}
\centering
\fbox{%
\begin{tabular}{cccccccc}\hline \\[-0.8em]
    &\multicolumn{7}{c}{Estimators} \\[-0.8em] \\ \\[-1em] \cline{3-6} \\[-0.8em]
\text{Mean (Standard Error)}  & \multicolumn{2}{c}{$\hat{\beta}^\ast_{\text{EM}}$} &  \multicolumn{2}{c}{$\hat{\beta}_{\text{EM}}$}&  \multicolumn{2}{c}{$\hat{\beta}^\ast_{\text{semi}}$}& $\hat{\beta}_{\text{semi}}$ \\[-0.8em] \\ \cline{1-1} \\[-0.8em]
n = 300  &1.03 (0.50)  &   &2.08 (0.47) &  & 2.09 (0.81) &  & 2.12 (0.80) \\
n = 500  &0.968 (0.38)  &   &2.03 (0.33) &  & 2.08 (0.69) &  & 2.08 (0.67) \\
n = 1000 & 0.951 (0.28)  &  &2.00 (0.24) &  & 1.99 (0.53) &  & 2.03 (0.49) \\\hline \hline
\\[-0.8em]
$|$\text{Bias$|$ (\% Bias)}  &  \\[-0.8em] \\ \cline{1-1} \\[-0.8em]
n = 300  & 0.97 ($48.5\%$)  &  & 0.08 ($4.00\%$) &  & 0.09 ($4.50\%$) &  & 0.12 ($6.00\%$) \\
n = 500  & 1.03 ($51.6\%$)  &  & 0.03 ($1.50\%$)  &  & 0.08 ($4.00\%$) &  & 0.08 ($4.00\%$) \\
n = 1000 & 1.05 ($52.5\%$)  &  & 0.00 ($0.00\%$)  &  & 0.01 ($0.50\%$) &  & 0.03 ($1.50\%$) \\\hline \hline
\\[-0.8em]
\text{Coverage of 95\% CI}  &  \\[-0.8em] \\ \cline{1-1} \\[-0.8em]
n = 300  &$64.5\%$  &  &$94.7\%$ &  & $95.0\%$ &  & $95.8\%$ \\
n = 500  &$35.5\%$  &  &$94.1\%$  &  & $93.9\%$ &  & $94.0\%$ \\
n = 1000 &$7.30\%$  &  &$95.4\%$  &  & $92.5\%$ &  & $93.7\%$ \\\hline \hline
\\[-0.8em]
\text{RMSE}  & \\[-0.8em] \\ \cline{1-1} \\[-0.8em]
n = 300  &1.09  &  &0.473  &  & 0.809 &  & 0.817 \\
n = 500  &1.10  &  &0.333  &  & 0.692 &  & 0.679 \\
n = 1000 &1.08  &  &0.239  &  & 0.532 &  & 0.493 \\\hline
\end{tabular}}
\end{table}

\begin{figure}
  \centering
  \subfigure[\small$\hat{\beta}^\ast_{\text{EM}}$: incorrectly specified U]{%
    \label{fig: binary U binary Y incorrect U EM}%
    \includegraphics[height = 7 cm, width = 0.5\linewidth]{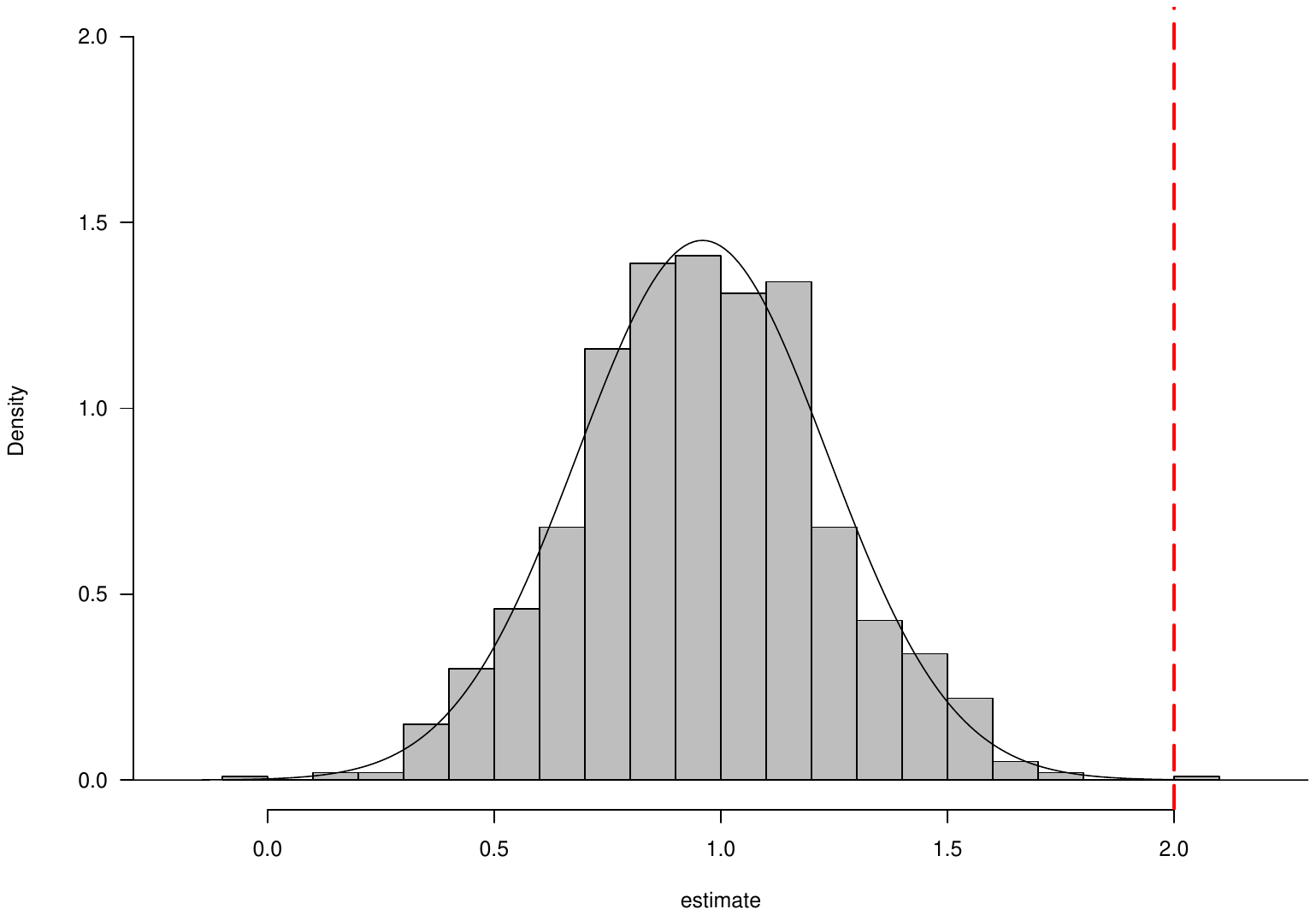}}%
  \subfigure[\small$\hat{\beta}_{\text{EM}}$: correctly specified U]{%
    \label{fig: binary U binary Y correct U EM}%
    \includegraphics[height = 7 cm, width = 0.5\linewidth]{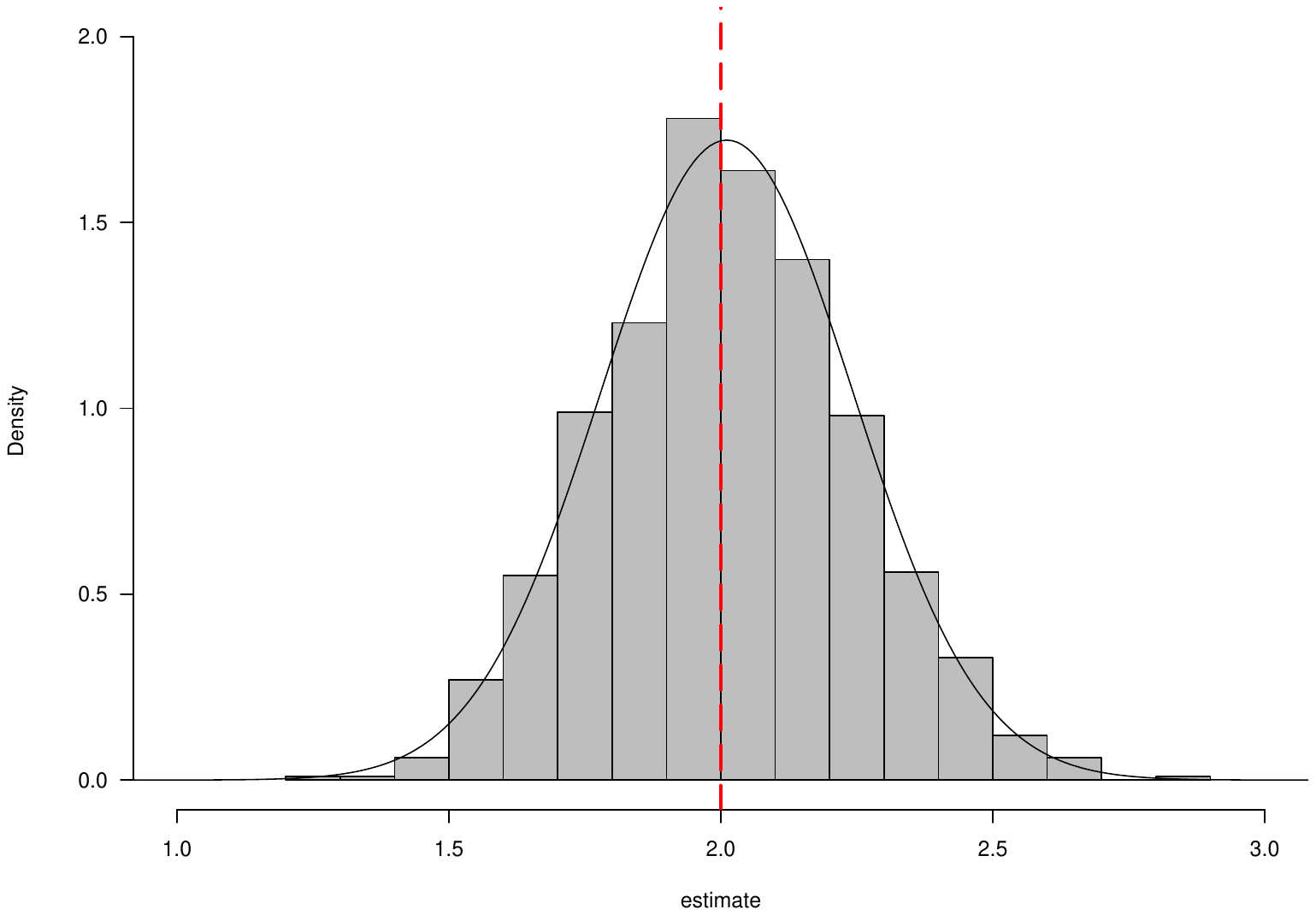}}
    \newline
  \centering
  \subfigure[\small $\hat{\beta}^\ast_{\text{semi}}$: incorrectly specified U]{%
     \label{fig: binary U binary Y incorrect U}%
    \includegraphics[height = 7 cm, width = 0.5\linewidth]{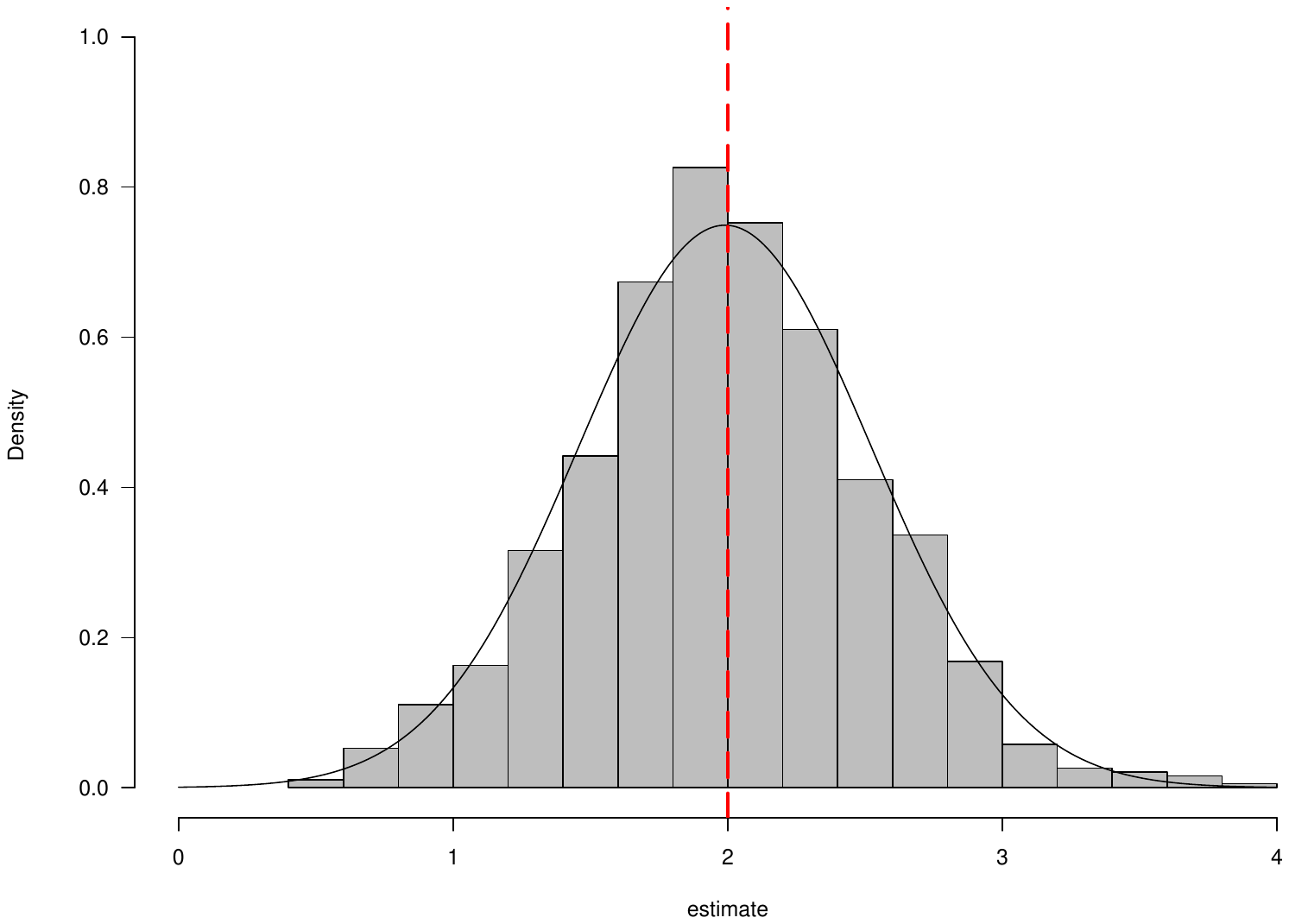}}%
  \subfigure[\small $\hat{\beta}_{\text{semi}}$: correctly specified U]{%
    \label{fig: binary U binary Y correct U}%
    \includegraphics[height = 7 cm, width = 0.5\linewidth]{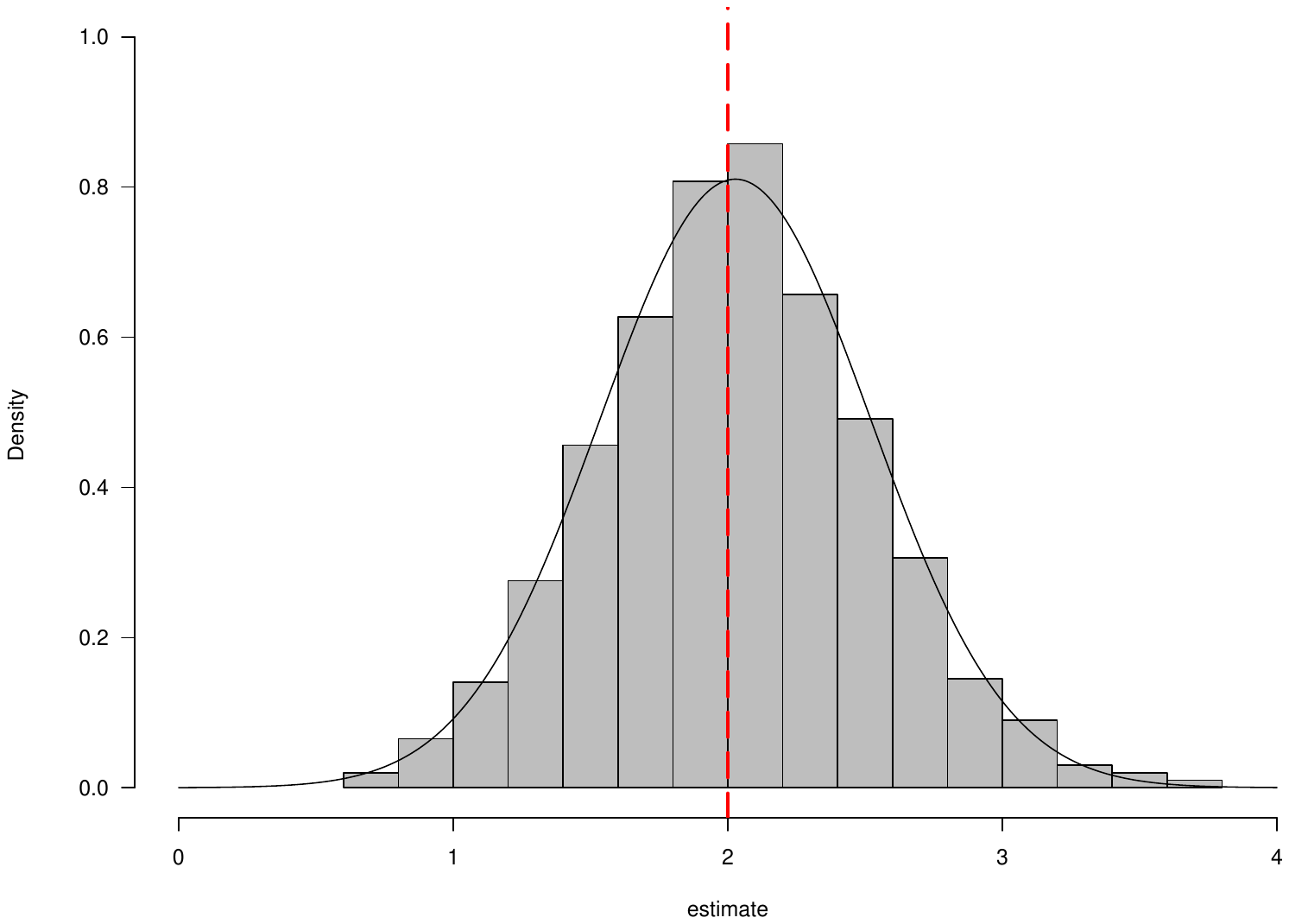}} 
  \caption{\small Four estimators when both $U$ and $Y$ are binary. Top two panels: maximum likelihood estimators via the EM algorithm. Bottom two panels: semiparametric estimators. True $\beta$ value is represented by a red vertical line in all panels.}
  \label{fig: binary U binary Y}
\end{figure}

Table 1 summarizes the Monte Carlo results of four estimators with $1000$ repetitions of experiments. The fully parametric specification is susceptible to bias from misspecification of the working model for $U$: $\hat{\beta}^\ast_{\text{EM}}$ is significantly biased with a $52.5\%$ of bias when $n = 1000$. On the other hand, our proposed semiparametric estimators $\hat{\beta}_{\text{semi}}$ and $\hat{\beta}^\ast_{\text{semi}}$ are always consistent (with a $0.50\%$ and $1.50\%$ of bias, respectively) and have approximately correct coverage, even when the specified distribution of $U$ is incorrect. In terms of efficiency, the semiparametric estimator $\hat{\beta}_{\text{semi}}$ (SE($\hat{\beta}_{\text{semi}}$) = 0.49 when $n = 1000$) with a correctly specified $U$ is more efficient than $\hat{\beta}^\ast_{\text{semi}}$ (SE($\hat{\beta}^\ast_{\text{semi}}$) = 0.53 when $n = 1000$), with a corresponding ARE($\hat{\beta}_{\text{semi}}$, $\hat{\beta}^\ast_{\text{semi}}$) = $1.08$. The maximum likelihood estimator $\hat{\beta}_{\text{EM}}$ with a correctly specified $U$ is the most efficient (SE($\hat{\beta}^\ast_{\text{semi}}$) = 0.24 when $n = 1000$), with a corresponding ARE($\hat{\beta}_{\text{EM}}$, $\hat{\beta}_{\text{semi}}$) = $2.21$. Figure \ref{fig: binary U binary Y} plots the Monte Carlo distributions of four estimators when $n = 1000$. Similar plots of the Monte Carlo distributions of semiparametric estimators for $n = 300$ and $n = 500$ can be found in the Supplementary Material E.1, and for a binary $U$ and a continuous $Y$ can be found in the Supplementary Material E.2.

\subsection{Continuous unmeasured confounder}
\label{subsec: simulation cont U}
In this section, we assess the performance of our proposed estimator when $U$ is continuous and the integral equation \eqref{eqn: integral equation} is approximated using the Tikhonov regularization and discretization as detailed in Supplementary Material B. We considered a data-generating process similar to Model \eqref{model: simulation model binary U} except that $U \sim \text{Beta}(2, 2)$ and $c_\gamma = c_\delta = 2$. Our working model for $U$ was a discrete distribution with equally-spaced support points on the unit interval with mesh size $h$. We constructed the proposed estimator for various combinations of sample size $n$ and mesh size $h$, and regularization parameter $\alpha = 0.1$. Table 2 summarizes the mean and standard error when the experiment is repeated $1000$ times. We also performed simulations for $n = 300$, $\alpha = 0.01$, and various mesh sizes. The results are similar to those with $\alpha = 0.1$.

\begin{figure}%
  \centering
  \subfigure[\small $h = 0.1$, $n = 300$]{%
    \label{fig: cont U n 300}%
    \includegraphics[width = 0.33\linewidth]{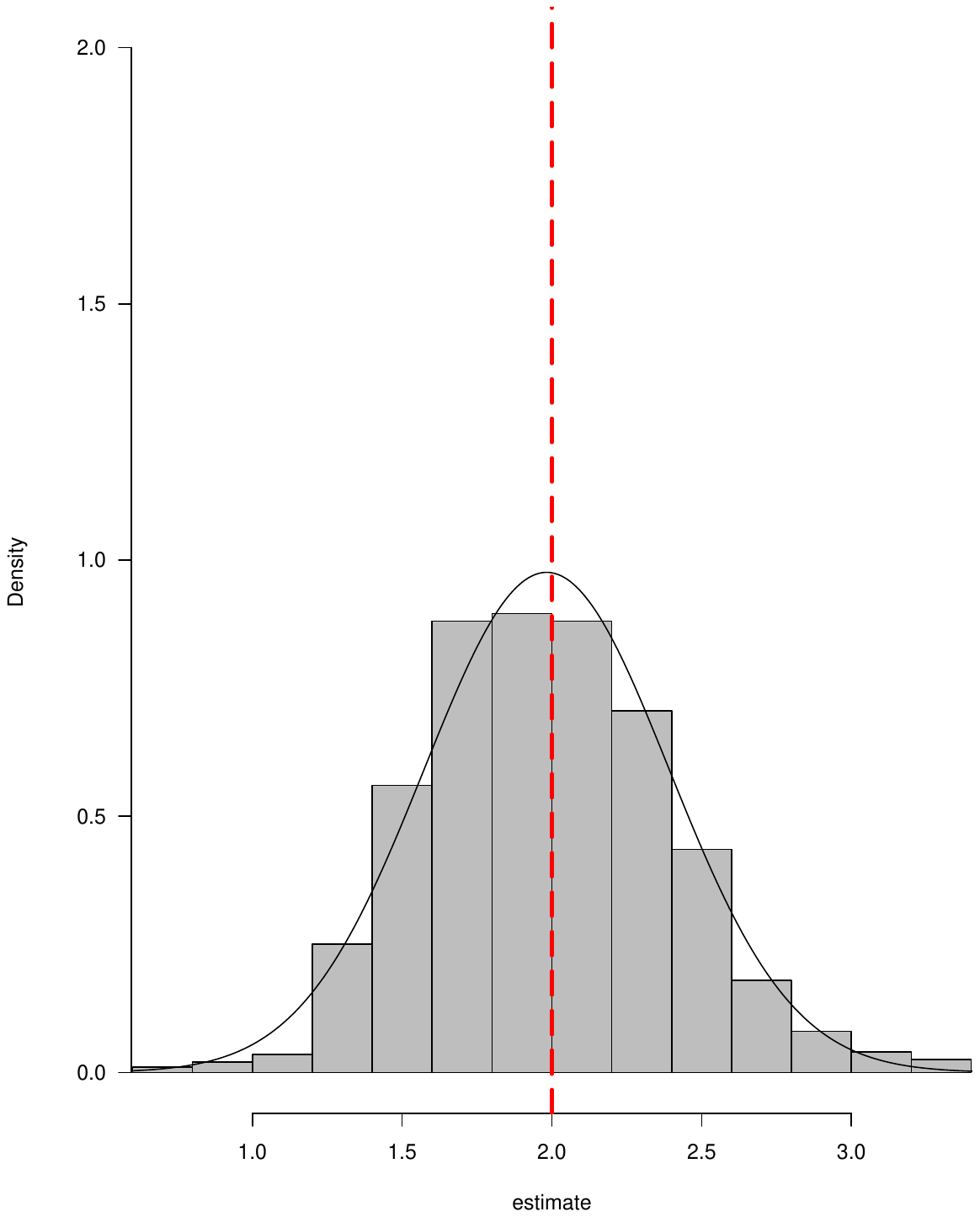}}
  \subfigure[\small $h = 0.1$, $n = 500$]{%
     \label{fig: cont U n 500}%
    \includegraphics[width = 0.33\linewidth]{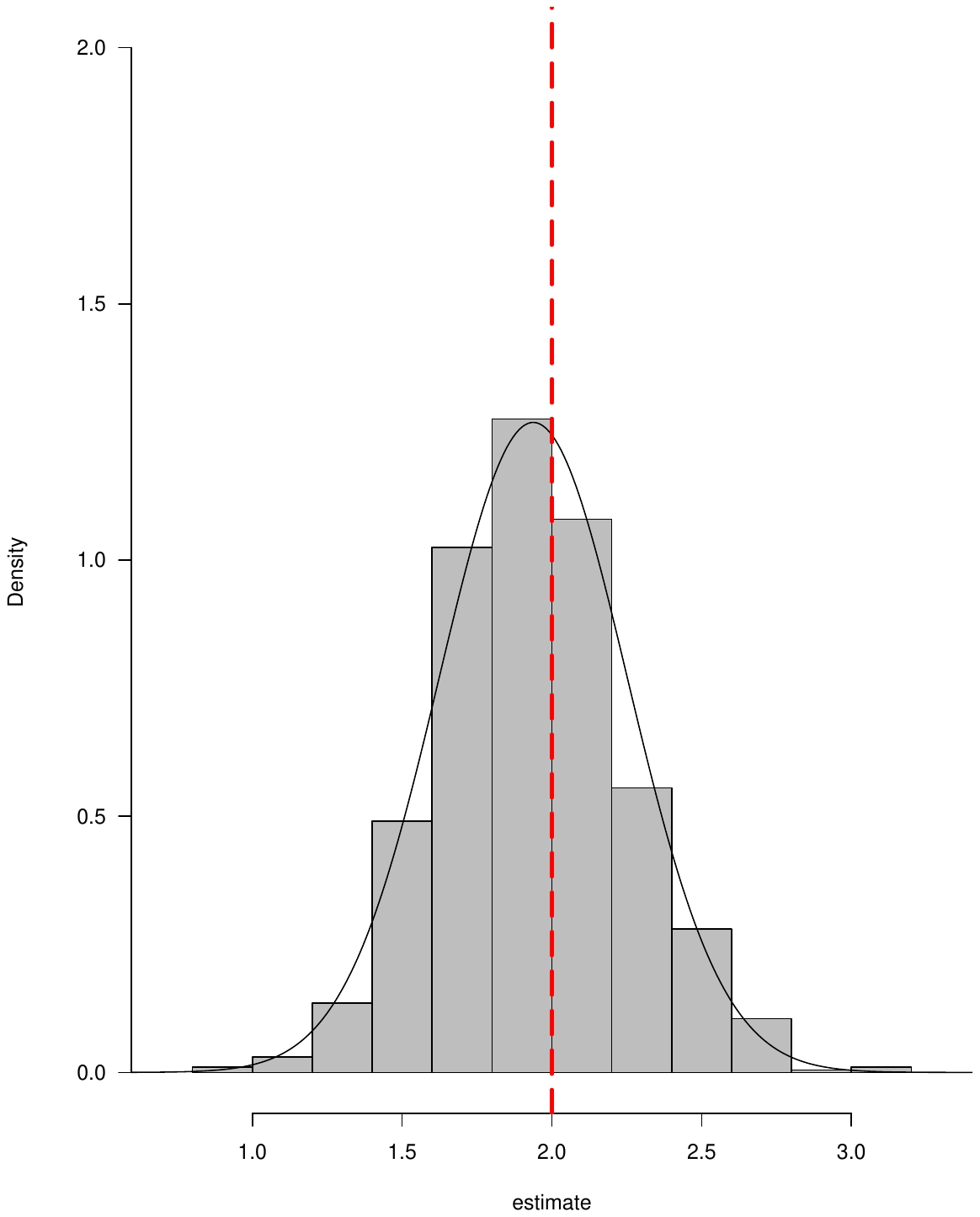}}%
  \subfigure[\small $h = 0.1$, $n = 1000$]{%
    \label{fig: cont U n 1000}%
    \includegraphics[width = 0.33\linewidth]{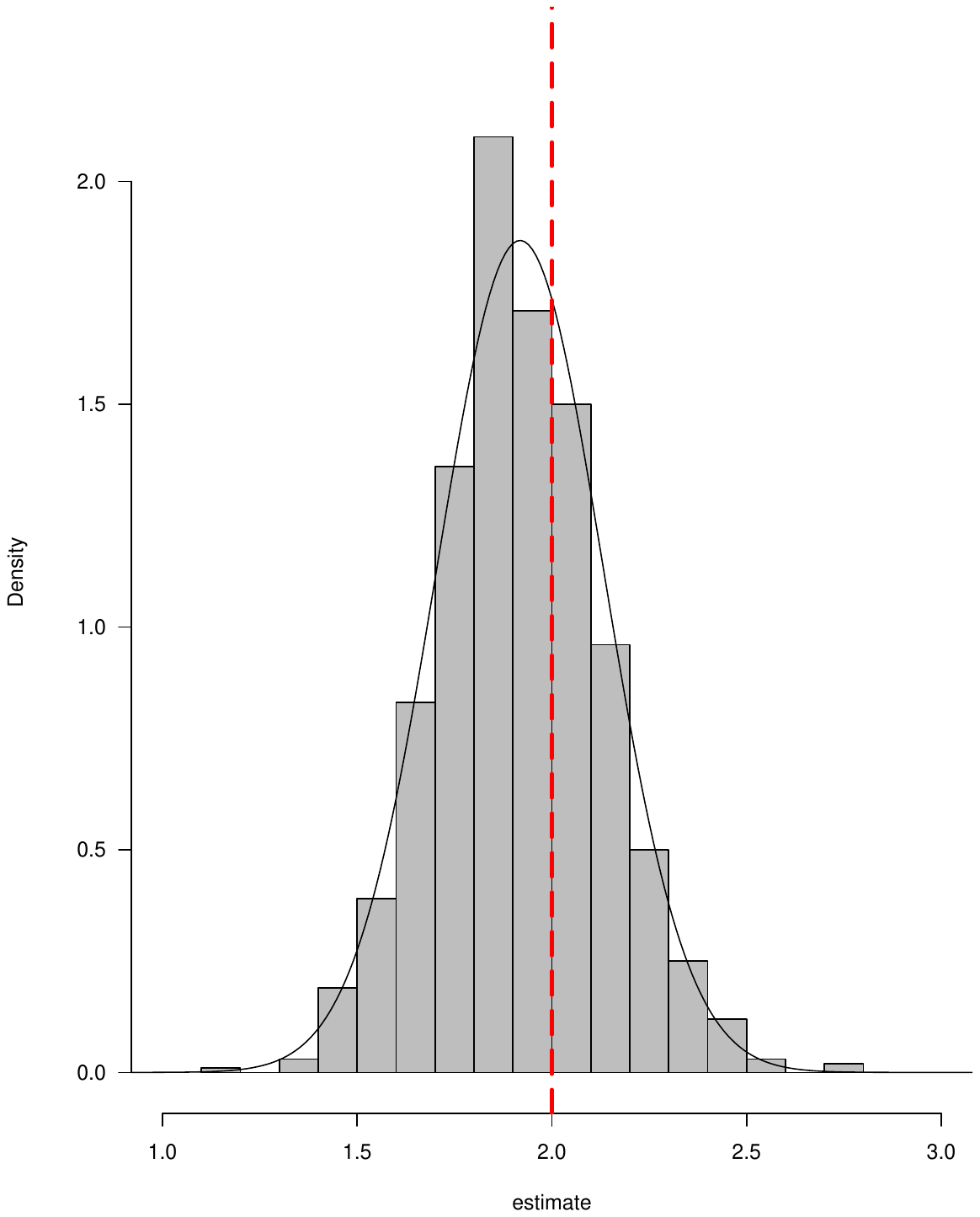}}%
    
  \caption{\small Semiparametric estimators with various degree of approximation when $U$ is continuous. True $\beta$ value is represented by a red vertical line in all three panels. Regularization parameter $\alpha = 0.1$.}
  \label{fig: three semi est continuous U}
\end{figure}

In each row, for a fixed sample size, we found that the estimator appeared to converge to the true mean as the mesh size decreased. Though our theory holds when $\alpha \rightarrow 0$, $h \rightarrow 0$, and $n \rightarrow \infty$, performance of the proposed estimator appeared favorable when mesh size is as small as $0.1$ for a sample size $n = 1000$. A larger sample size $n$ often calls for a smaller mesh size $h$. In practice, practitioners could gradually decrease the mesh size $h$ up to a point when the estimator stabilizes. Computation time scales roughly as $O(n/h^3)$, where $O(n)$ comes from solving the integral equation for every data point, and $O(1/h^3)$ comes from inverting a $\ceil{1/h} \times \ceil{1/h}$ matrix. Replicating simulations $1000$ times takes $40$ minutes when $n = 500$ and $h = 0.2$, and roughly six hours when $n = 1000$ and $h = 0.1$ on a $64-$node cluster when executed in the programming language $\textsf{R}$. Figure \ref{fig: three semi est continuous U} further plots the Monte Carlo distributions of proposed estimator when $h = 0.1$, and $n = 300$, $n = 500$, and $n = 1000$.

\begin{table}
\caption{Monte Carlo results of the proposed estimator $\hat{\beta}_{\text{semi}}$ for various sample size $n$ and mesh size $h$: mean, standard error, bias, percentage of bias, coverage, and RMSE. Regularization parameter $\alpha = 0.1$. True $\beta$ equals $2.0$.}
\label{tb: continuous U indep X}
\centering
\fbox{%
\begin{tabular}{cccccccc}\hline \\[-0.8em]
    &\multicolumn{7}{c}{Mesh size h} \\[-0.9em] \\ \\[-1em] \cline{3-6} \\[-0.8em]
\text{Mean (Standard Error)}  & \multicolumn{2}{c}{0.5} &  \multicolumn{2}{c}{0.25}&  \multicolumn{2}{c}{0.2}& 0.1 \\[-0.8em] \\ \cline{1-1} \\[-0.8em]
n = 300  &  1.87 (0.42) &  &1.91 (0.40) &  & 1.94 (0.40) &  & 1.96 (0.40) \\
n = 500  &  1.82 (0.31) &  &1.88 (0.31) &  & 1.91 (0.31) &  & 1.95 (0.31) \\
n = 1000 &  1.81 (0.22) &  &1.87 (0.22) &  & 1.91 (0.21) &  & 1.92 (0.22) \\\hline \hline
\\[-0.8em]
$|$\text{Bias$|$ (\% Bias)}  &  \\[-0.8em] \\ \cline{1-1} \\[-0.8em]
n = 300  &  0.13 ($6.50\%$) &  & 0.09 ($4.50\%$) &  & 0.06 ($3.00\%$) &  & 0.04 ($2.00\%$) \\
n = 500  &  0.18 ($9.00\%$) &  & 0.12 ($6.00\%$) &  & 0.09 ($4.50\%$) &  & 0.05 ($2.50\%$) \\
n = 1000 &  0.19 ($9.50\%$) &  & 0.13 ($6.50\%$) &  & 0.09 ($4.50\%$) &  & 0.08 ($4.00\%$) \\\hline \hline
\\[-0.8em]
\text{Coverage of 95\% CI}  &  \\[-0.8em] \\ \cline{1-1} \\[-0.8em]
n = 300  &  $91.3\%$ &  & $93.6\%$ &  & $94.0\%$ &  & $94.0\%$ \\
n = 500  &  $88.3\%$ &  & $91.9\%$ &  & $92.6\%$ &  & $94.3\%$ \\
n = 1000 & $83.3\%$ &  & $89.5\%$ &  & $92.7\%$ &  & $92.4\%$ \\\hline \hline
\\[-0.8em]
\text{RMSE}  & \\[-0.8em] \\ \cline{1-1} \\[-0.8em]
n = 300  & 0.434 &  & 0.413 &  & 0.405 &  & 0.400 \\
n = 500  & 0.360 &  & 0.326 &  & 0.318 &  & 0.314 \\
n = 1000 & 0.290 &  & 0.251 &  & 0.233 &  & 0.229 \\\hline
\end{tabular}}
\end{table}

\subsection{Dependent continuous unmeasured confounder}
\label{subsec: dependent cont U}
We consider the case where $U$ is allowed to depend on $(X_1, X_2)$ in this section. We considered a data-generating process similar to Model \eqref{model: simulation model binary U} except that we let $U = X_1 + \text{Beta}(2, 2)$. In this setting, we used the same working model as before and performed $1000$ simulations at three sample sizes ($n = 200$, $300$, and $500$) and two different mesh sizes ($h = 0.2$ and $0.1$). Next, we let $U = X_1 + \text{Normal}(0, 0.1)$ and used $U \sim \text{Unif}[-0.4, 0.4]$ as our working model. Again, we repeated the simulation $1000$ times at two different mesh sizes ($h = 0.1$ and $0.05$) and three different sample sizes ($n = 200$, $300$, and $500$). Table 3 summarizes the Monte Carlo results. Again, we observed that the estimator appeared to converge to the true value as mesh size decreased, and the coverage of the constructed confidence intervals appeared to approximately achieve their nominal levels. 

\begin{table}
\caption{Monte Carlo results of the proposed estimator $\hat{\beta}_{\text{semi}}$ when $U_i = X_{i,1} + \epsilon_i$ and for various sample size $n$ and mesh size $h$: mean, standard error, bias, percentage of bias, coverage, and RMSE. Regularization parameter $\alpha = 0.1$. True $\beta$ equals $2.0$.}
\label{tb: continuous U dep X}
\centering
\fbox{%
\begin{tabular}{ccccccc}\hline \\[-0.8em]
    &\multicolumn{2}{c}{$\epsilon_i = \text{Beta}(2, 2)$} &&\multicolumn{2}{c}{$\epsilon_i = \text{Normal}(0, 0.1)$} \\[-0.9em] \\ \\[-1em] \cline{2-3} \cline{5-6} \\[-0.8em]
\text{Mean (Standard Error)}  & h = 0.2 &h = 0.1 && h = 0.1& h = 0.05 \\[-0.9em] \\ \cline{1-1} \\[-0.9em]
n = 200  & 1.96 (0.61)  &1.99(0.64) && 1.94 (0.53) & 1.97 (0.53)  \\
n = 300  &  1.97 (0.53) &1.98 (0.51) && 1.93 (0.42) & 1.94 (0.43)  \\
n = 500  &  1.93 (0.38)  &1.94 (0.37) &&1.91 (0.33)  & 1.93 (0.34) \\ \hline \hline
\\[-0.8em]
\text{$|$\text{Bias}$|$ (\% Bias)}   \\[-0.9em] \\ \cline{1-1} \\[-0.9em]
n = 200  & 0.04 ($2.00\%$) & 0.01 ($0.50\%$) && 0.06 ($3.00\%$) & 0.03 ($1.50\%$)  \\
n = 300  & 0.03 ($1.50\%$) & 0.02 ($1.00\%$) && 0.07 ($3.50\%$) & 0.06 ($3.00\%$)  \\
n = 500  & 0.07 ($3.50\%$) & 0.06 ($3.00\%$) && 0.09 ($4.50\%$) & 0.07 ($3.50\%$)\\ \hline \hline
\\[-0.8em]
\text{Coverage of 95\% CI}   \\[-0.9em] \\ \cline{1-1} \\[-0.9em]
n = 200  & $94.9\%$  & $95.1\%$ && $93.8\%$ & $95.4\%$  \\
n = 300  & $93.4\%$ & $93.8\%$ && $95.0\%$ & $93.3\%$  \\
n = 500  & $93.8\%$  & $94.3\%$ && $93.1\%$ & $92.0\%$\\ \hline \hline
\\[-0.8em]
\text{RMSE}   \\[-0.9em] \\ \cline{1-1} \\[-0.9em]
n = 200  & 0.608  & 0.637 && 0.531 & 0.534  \\
n = 300  & 0.529 & 0.506 && 0.427 & 0.436  \\
n = 500  & 0.383  & 0.378 && 0.341 & 0.345 \\ \hline
\end{tabular}}
\end{table}

\section{REPORTING A SENSITIVITY ANALYSIS}
\label{sec: one-param SA}
\subsection{One-parameter vs. two-parameter sensitivity analysis}
\label{subsec: one and two parameter}
Semiparametric model $\mathcal{M}$ consists of a rich collection of laws. It allows empirical researchers to specify two sensitivity parameters: one controlling the strength of association between $U$ and $Y$, and the other between $U$ and $Z$. Such a two-parameter sensitivity analysis has a long history in the causal inference literature. The first sensitivity analysis carried out by \citet{Cornfield1959} for observational studies of cigarette smoking as a cause of lung cancer adopted this ``two-parameter'' paradigm. A lot of subsequent methodological development (e.g., those referenced in Section \ref{subsec: review added variable models}) fall into this category. 


The most comprehensive output of a two-parameter sensitivity analysis may be a graph with x-axis being one sensitivity parameter (e.g. $c_\gamma$) and y-axis the other (e.g. $c_\delta$); see, e.g., \citet{Rosenbaum2009, Griffin2013,Hsu2013,ding2016sensitivity,zhang2018calibrated}, among others. In practice, however, a plot may be too cumbersome for some empirical studies where many aspects of an analysis need to be examined but the space is very much limited. One natural question is: Is it \emph{necessary} to correctly specify both the outcome model $f(Y \mid Z,\mathbf{X},U)$ and the propensity score model $f(Z \mid \mathbf{X}, U)$? Can we further relax the modeling assumption on one of the two models, and summarize the sensitivity analysis using only one sensitivity parameter $c_\delta$ (association between $U$ and $Y$) or $c_\gamma$ (association between $U$ and $Z$)? Unfortunately, the answer to this question is negative, as is illustrated in the following toy example.

\begin{example}[Non-identifiability of one-parameter sensitivity analysis]\rm
Consider a simple data-generating process as follows: $P(U = 1) = p$, $P(Z = 1) = \text{expit}(\lambda_0 + c_\gamma U)$, and $P(Y = 1) = \text{expit}(\beta_0 + \beta_1 Z + c_\delta U)$. Let $\theta = (p, \lambda_0, c_\gamma, \beta_0, \beta_1)$ and fix $c_\delta = 1$ as our sensitivity parameter. One can easily check that $\theta_1 = (1, -0.5, 0.5, -0.5, -0.5)$ and $\theta_2 = (1, -1, 1, -0.5, -0.5)$ yield the same observed data likelihood: $P(Y = 1, Z = 1) = P(Y = 1, Z = 0) = P(Y = 0, Z = 1) = P(Y = 0, Z = 0) = 1/4$. 
\label{example: non-identifiability}
\end{example}

\citet{rosenbaum1987sensitivity, rosenbaum1989sensitivity} considered an alternative, one-parameter analysis which is a limiting case of the two-parameter analysis where the association between $U$ and $Z$ is held fixed and the association between $U$ and $Y$ goes to infinity. Such a one-parameter analysis is called a \emph{primal sensitivity analysis}, and the parallel limiting case where the association between $U$ and $Z$ goes to infinity is called a \emph{dual sensitivity analysis}. Our proposed method also works in harmony with this primal and dual framework: one may construct an estimator of $\beta$ with $c_\delta$ fixed at a very large value and $c_\gamma$ varying in a reasonable range, or vice versa. 

\subsection{Tipping point analysis vs. uniformly-valid confidence band}
Thus far, we have been making inference in a ``pointwise'' fashion, and outputting an estimate $\hat{\beta}$ of $\beta$ for fixed $(c_\delta, c_\gamma)$ values. This is the most common practice in sensitivity analysis literature, and is justified as the quantity of interest in empirical studies is often the tipping point pair $(c_\delta^\ast, c_\gamma^\ast)$, defined as the minimum strength of unmeasured confounding needed to explain away the observed treatment effect. In matched observational studies, tipping point sensitivity parameter is referred to as \emph{sensitivity value} (\citealp{zhao2019sensitivity}). Therefore, reporting confidence intervals corresponding to different $(c_\delta, c_\gamma)$ values can be thought of as an exercise searching for such a tipping point pair.

Alternatively, researchers can formally take into account uncertainty in sensitivity parameters $(c_\delta, c_\gamma)$ by constructing a confidence band of $\beta$ for sensitivity parameters falling in a feasible sensitivity parameters region $\Delta \times \Gamma$. For instance, one may specify $c_\delta \in [0, \Tilde{\delta}] = \Delta$ and $c_\gamma \in [0, \Tilde{\gamma}] = \Gamma$ for some chosen $\Tilde{\delta}$ and $\Tilde{\gamma}$. In Supplementary Material D.1, we describe how to construct such a confidence band using a version of the multiplier bootstrap.

\section{INTERPRETING THE SENSITIVITY ANALYSIS}
\label{sec: interpret}
\subsection{Two perspectives of $U$}
\label{subsec: two perspectives of U}
Applying and interpreting a sensitivity analysis critically depends on one's perspective of the unmeasured confounder $U$. There are at least two perspectives of $U$ that are relevant. In some cases, researchers may have in mind a specific candidate unmeasured confounder; for instance, a genetic variant when Sir Ronald Fisher challenged the causal interpretation of the association between smoking and lung cancer (\citealp{fisher1958cancer}). In other cases, researchers could use a scalar $U$ to represent all residual unmeasured confounding by defining $U \in [0, 1]$ to be the following quantity:
\begin{equation}
\label{eqn: U representing residual confounding}
    U = P(Z = 1 \mid \mathbf{X}, Y(1), Y(0)),
\end{equation}
so that the treatment assignment $Z$ is ignorable given $(\mathbf{X}, U)$, and is ignorable given $\mathbf{X}$ only when $U$ equals the propensity score (\citealp{rosenbaum2017adaptive}). If the empirical researcher is informed of the distribution of the candidate unmeasured confounder in the population under consideration, then one may directly leverage the model as in \citet{Rosenbaum1983} and  \citet{Imbens2003}. On the other hand, if the distribution of the unmeasured confounder is not known with confidence or there are multiple potential sources of unmeasured confounding, then the second perspective that treats $U$ as representing an aggregate of all possible residual, unmeasured confounding may be more favorable, and our proposed method is suitable for this case because our method naturally specifies the support of $U$ as being the unit interval without specifying its distribution.

\subsection{Interpretation}
\label{subsec: interpretation}

Inspired by \cite{Cornfield1959}, \cite{Gastwirth1998}, and \cite{Rosenbaum2002a}, we focus on two subjects with the same observed covariates. Consider a logistic regression relating the treatment assignment $Z$ to observed covariates $\mathbf{X}$ and the unmeasured confounder $U$ for subject $j$: \[
\log \frac{\pi_j}{1 - \pi_j} = \mathbf{\kappa}^T \mathbf{x} + c_\gamma u_j,
\] where $\pi_j = P(Z_j = 1 \mid \mathbf{x}, u_j)$ and $u_j$ is the unmeasured confounder associated with subject $j$. Consider another subject $k$ with the same observed covariates, but a possibly different unmeasured confounder $u_k$. As in \cite{Rosenbaum2002a}, the odds ratio of receiving treatment for two subjects with the same observed covariates is \[
\text{OR} = \frac{\pi_j(1-\pi_j)}{\pi_k/(1-\pi_k)} = \exp\{c_\gamma(u_j - u_k)\}.
\]

By treating $U$ as the aggregate of residual confounding as in \eqref{eqn: U representing residual confounding} so that $U \in [0, 1]$, $\text{OR}$ is bounded between $\exp(-c_\gamma)$ and $\exp(c_\gamma)$ and we can make the following statement: 
\begin{quote}
    Two subjects with the same observed covariates could differ in their odds of receiving the treatment, due to the unmeasured confounder, by \emph{at most} a factor of $\exp(c_\gamma)$.
\end{quote}

The interpretation of the other sensitivity parameter $c_\delta$ is more nuanced: it depends on the effect measure and the particular outcome regression model the practitioner chooses to fit. A general recipe is to follow \citet{Rosenbaum2002a} and think of how the outcome would systematically differ for subjects with the same observed covariates and receiving the same treatment, due to the unmeasured confounding. For instance, when the outcome is binary and a logistic regression model is fit to relate the binary outcome, the treatment, the observed covariates, and the unmeasured confounder, as in the running example, then we have a similar interpretation for $c_\delta$ as for $c_\gamma$: 
\begin{quote}
    Two subjects with the same observed covariates and receiving the same treatment could differ in their odds of receiving the outcome, due to the unmeasured confounder, by \emph{at most} a factor of $\exp(c_\delta)$.
\end{quote}
When the outcome is continuous, a popular choice is to fit a linear regression for the outcome, as in \citet{Imbens2003}: \[
Y_i(z) \mid \mathbf{X}_i, U_i \sim \text{Normal}(\beta z + \mathbf{\lambda}^T \mathbf{X}_i + c_\delta U_i, \sigma^2).
\]
One quantity of interest in this scenario is $c_\delta/\sigma$ and a proper interpretation of the sensitivity analysis results is the following:
\begin{quote}
    Two subjects with the same observed covariates and the same treatment may vary in their response, in the mean scale, by \emph{at most} $c_\delta/\sigma$ standard deviations. 
\end{quote}

\subsection{Comparison to Rosenbaum bounds}
\label{subsec: comparison to rosenbaum bounds}
Our method extends the work by \citet{Rosenbaum1983} and \citet{Imbens2003}; it can also be viewed as a generalization of Rosenbaum bounds in a matched observational study (\citealp{Rosenbaum2002a, DiPrete2004}). \citet{Rosenbaum2002a}'s analytical framework concerns about finite-sample inference of Fisher's sharp null hypothesis, and views the collection of unmeasured confounders of all study units $\mathbf{u}$, as fixed attributes of the sample. Fix a sensitivity parameter $c_\gamma$ and hence the maximum odds ratio among all matched pairs or sets, \citet{Rosenbaum2002a}'s sensitivity analysis proceeds by calculating the bound on the tail probability of a test statistic under the sharp null hypothesis over nuisance parameters $\mathbf{u}$ supported on a nuisance parameter space $\mathcal{U}$; for instance, in a matched pair design with $I$ pairs and $2I$ study units, the nuisance parameter space $\mathcal{U} = [0, 1]^{2I}$, and the bounding $p$-value corresponding to a fixed $c_\gamma$ value is valid for any possible realization of $\mathbf{u} \in \mathcal{U}$. On the other hand, we adopt a superpopulation perspective, view the unmeasured confounder $U$ as a random variable in the population, and output a valid $p$-value and confidence interval of the treatment effect for any distribution of $U$. To summarize, both \citet{Rosenbaum2002a} and our method specifies the extent of maximum deviation of odds ratio from randomization, \citet{Rosenbaum2002a}'s method is valid for any realizations of $\mathbf{u}$ subject to the maximum deviation constraint, while our method is valid for any distribution on the random variable $U$ subject to the same maximum deviation constraint.


\section{WAR AND POLITICAL PARTICIPATION REVISITED}
\label{sec: application}
We are now ready to investigate the sensitivity of the war and political participation study in Uganda to unmeasured confounding using the proposed method. In our analysis, we controlled for father's education, mother's education, family size, and whether parents died before abduction. The treatment is binary, equal to $1$ if the subject had been abducted and $0$ otherwise, and the outcome of interest is whether the subject voted in the $2005$ referendum in Uganda. As the database contains missing data, we performed a multiple imputation with $5$ replicates using the \textsf{mice} package (\citealp{R_mice}) in $\mathsf{R}$ with default settings, and combined estimates using Rubin's rules (\citealp{Rubin1987}).

\begin{figure}
    \centering
    \includegraphics[width=0.8\textwidth]{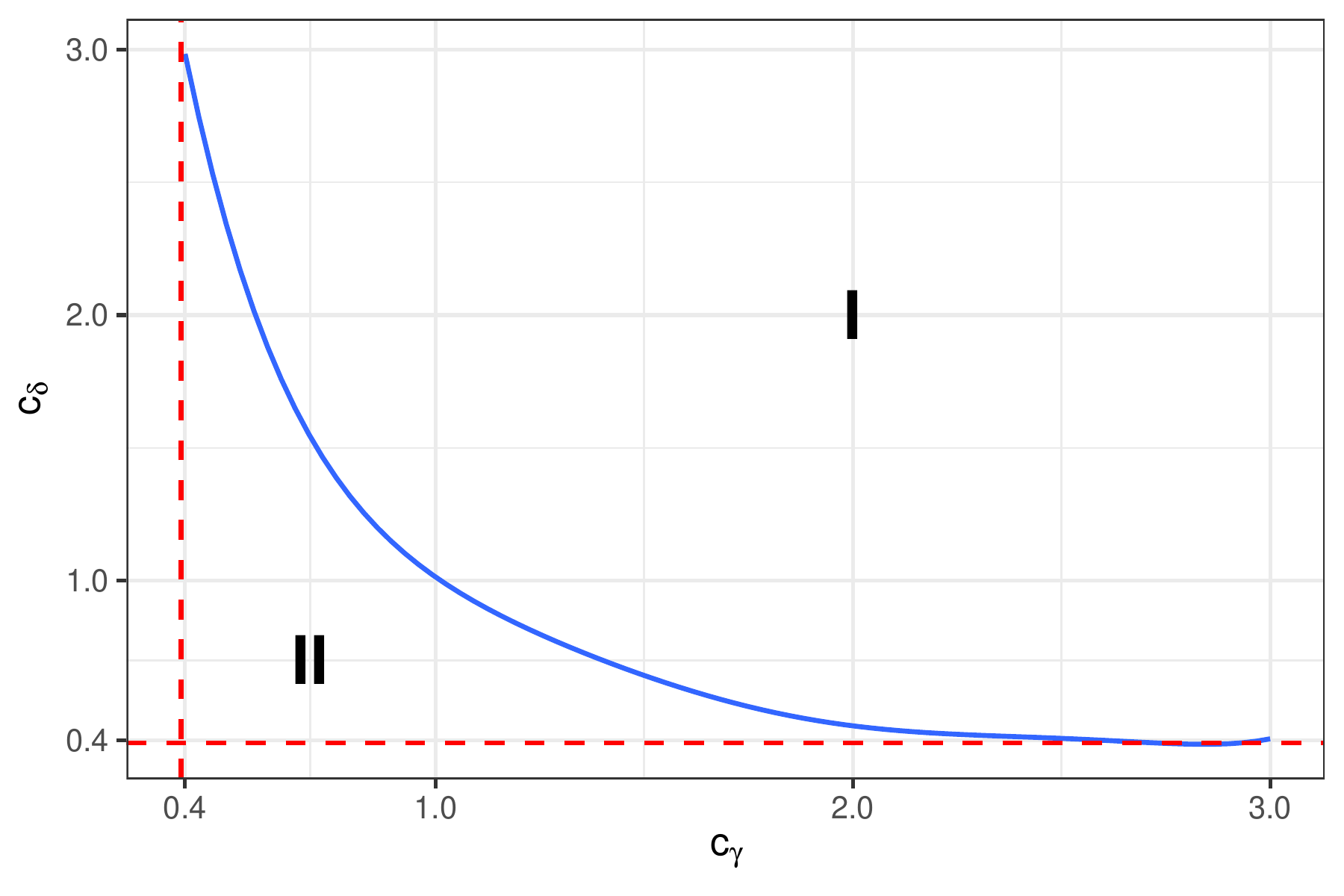}
    \caption{Region to the left of the solid curve (Region II) contains sensitivity parameter pairs $(c_\delta, c_\gamma)$ for which the $95\%$ confidence intervals do not contain $0$. Two dotted lines can be interpreted as corresponding to a primal ($c_\gamma = 0.4$) and dual ($c_\delta = 0.4$) sensitivity analysis, respectively.}
    \label{fig: real data contour}
\end{figure}

We view the hypothesized unmeasured confounder $U$ as the aggregate of many potential sources of bias. As discussed in Section \ref{subsec: two perspectives of U}, we may stipulate $U \in [0, 1]$ without loss of generality. We related the treatment assignment $Z$ to observed covariates $\mathbf{X}$ and $U$ using a logistic regression, related the outcome $Y$ to $\mathbf{X}$, $Z$, and $U$ using a logistic regression with a constant additive effect, as \cite{Blattman2010} did in their original analysis, and put a uniform distribution on $U$ as a working model when constructing the semiparametric estimator. For a fixed $(c_\delta, c_\gamma)$ combination, our procedure entails the following steps:
\begin{enumerate}
    \item Approximate the solution to equation \eqref{eqn: integral equation} at each $\mathbf{X} = \mathbf{X}_i$;
    \item Calculate the efficient score according to \eqref{eqn: efficient observed data score} at each data point;
    \item Set up the estimating equations and obtain estimators of model parameters including an estimator of the treatment effect $\hat{\beta}$;
    \item Compute the robust sandwich estimator of the variance of $\hat{\beta}$;
    \item Construct a $95\%$ confidence interval of $\beta$.
\end{enumerate}
We repeat the above procedure at different $(c_\delta, c_\gamma)$ combinations and summarize the results in Figure \ref{fig: real data contour}. Region to the left of the solid curve contains sensitivity parameter pairs $(c_\delta, c_\gamma)$ for which $95\%$ confidence intervals contain $0$. All tipping point sensitivity parameters $(c_\delta, c_\gamma)$ are captured by the solid contour curve and they admit interpretation as outlined in Section \ref{subsec: interpretation}. For instance, combination $(c_\delta, c_\gamma) = (1.0, 1.0)$ on the curve has the following interpretation: The observed treatment effect remains significant at the $0.05$ level if the following two conditions are satisfied simultaneously:
\begin{enumerate}
    \item Two subjects with the same observed covariates differ in their odds of being abducted by the LRA, due to unmeasured confounding, by no more than a factor of $\exp(1.0) = 2.72$.
    \item Two subjects with the same observed covariates and treatment status differ in their odds of voting in the 2005 referendum, due to unmeasured confounding, by no more than a factor of $\exp(1.0) = 2.72$.
\end{enumerate}

Figure \ref{fig: real data contour} is arguably the most comprehensive sensitivity analysis output from a tipping point analysis perspective. Readers who are interested in a uniformly valid confidence band that formally accounts for the uncertainty in $(c_\delta, c_\gamma)$ values may refer to the Supplementary Material D.2 for such a result. 

As discussed in Section \ref{subsec: one and two parameter}, a two-parameter sensitivity analysis simultaneously places constraints on $Z-U$ and $Y-U$ associations; alternatively, researchers could elect to report a primal sensitivity analysis (\citealp{rosenbaum1989sensitivity}) by sending $c_\delta$ to infinity and only reporting the $Z-U$ association as captured by $c_\gamma$. The vertical dashed line in Figure \ref{fig: real data contour} corresponds to such a primal sensitivity analysis, and the horizontal dashed line a parallel, dual sensitivity analysis. The primal sensitivity analysis can be interpreted in the following way: If the unmeasured confounding does not increase the odds of being abducted by the LRA by a factor of $\exp(0.4) = 1.49$, then the effect would still be significant at the $0.05$ level no matter how strongly the unmeasured confounding is associated with voting in the 2005 referendum.

\section{DISCUSSION}
\label{sec: conclude}
In this paper, we proposed a novel semiparametric approach to model-based sensitivity analysis. We showed how to relax the parametric assumption often imposed on the unmeasured confounder and still draw valid inference under the popular model-based sensitivity analysis framework proposed by \citet{Rosenbaum1983} and extended by \citet{Imbens2003}. There are at least three advantages of relaxing this piece of assumption. First, the class of models under consideration is more flexible and largely reduces what \citet{franks2019flexible} called \emph{observable implications}. Second, it facilitates thinking about the robustness of a sensitivity analysis: different parametric assumptions on $U$ might yield different conclusions and it is ideal that a sensitivity analysis can be robust to different specifications of $U$. Moreover, our approach works seamlessly with any primary analysis that models $\mathbb{E}[Y \mid Z, \mathbf{X}]$ parametrically, which is still a widely used strategy in the empirical causal inference literature. To make the outcome model more robust, one may first perform a nonparametric preprocessing step, say via statistical matching, and do regression adjustment within each matched set by including matched-set specific fixed effects (\citealp{Rubin1979,ho2007matching,zhang2018calibrated}). However, solving estimating equations with a large number of parameters (matched-set fixed effect) can be challenging. While we only investigate the canonical setting where we have a point exposure and one outcome of interest, our framework could be potentially extended to many other settings: e.g., the setting where exposure and covariates are all time-varying, and the setting of instrumental variable analysis where there is still concern about residual IV-outcome confounding.

\clearpage
\pagenumbering{arabic}
\begin{center}
    {\LARGE\bf Online Supplementary Materials for ``A Semiparametric Approach to Model-Based Sensitivity Analysis in Observational Studies" by Bo Zhang and Eric J. Tchetgen Tchetgen}
\end{center}
  \medskip

\begin{center}
{\large\bf Supplementary Material A: Geometry and Observed Data Efficient Influence Functions}
\end{center}
\bigskip

We discuss the geometry of the Hilbert space associated with the full data and the observed data in our problem. We then leverage the geometry to derive the efficient influence function which motivates the estimating equation. 

Recall that the full data consist of $\mathcal{D} = \{D_i = (\mathbf{X}_i, U_i, Z_i, Y_i), ~i = 1,2,...,n\}$. The full data nuisance tangent space is given by $\Lambda^F = \Lambda_{1}^F \oplus \Lambda_{2}^F$, where
\begin{gather*}
\Lambda_{1}^F = \{a_1(\mathbf{X}): \mathbb{E}[a_1(\mathbf{X})] = 0\} \cap L^2, \\
\Lambda_{2}^F = \{a_2(U, \mathbf{X}): \mathbb{E}[a_2(U, \mathbf{X}) \mid \mathbf{X}] = 0\} \cap L^2.
\end{gather*}
Because $U$ is not observed, the observed data consist of $\mathcal{O} = \{O_i = (\mathbf{X}_i, Z_i, Y_i), i = 1,2,...,n\}$ and by standard semiparametric theory, the observed data nuisance tangent space $\Lambda$ is the projection of the full data nuisance tangent space onto the observed data (\citealp{Bickel1993efficient,Robins1994}), i.e., $\Lambda = \Lambda_{1} \oplus \Lambda_{2}$, where
\begin{gather*}
\Lambda_{1} = \mathbb{E}[\Lambda_{1}^F \mid \mathbf{X}, Z, Y] = \Lambda_{1}^F,\\
\Lambda_{2} = \mathbb{E}[\Lambda_{2}^F \mid \mathbf{X}, Z, Y] = \{\mathbb{E}[a_2(U, \mathbf{X}) \mid \mathbf{X}, Z, Y]: \mathbb{E}[a_2(U, \mathbf{X}) \mid \mathbf{X}] = 0\}.
\end{gather*}

Let $\theta = (\lambda, \beta, \kappa)$ denote the finite dimensional parameter of interest and $S_\theta(\mathbf{X}, U, Z, Y)$ the full data score of $\theta$. Observed data scores are then obtained by projecting full data scores onto the observed data (\citealp{Bickel1993efficient,Robins1994}): \begin{equation*}
\text{Observed data score} = S_\theta(\mathbf{X}, Z, Y)
= ~\mathbb{E}[S_\theta(\mathbf{X}, U, Z, Y) \mid \mathbf{X}, Z, Y].
\end{equation*}

The key step in deriving the efficient observed data influence function is to project the observed data score onto the ortho-complement to the observed data nuisance tangent space. Theorem \ref{thm: projection in H} provides an expression for the ortho-complement to the nuisance tangent space and derives the observed data efficient score $S_\text{eff}(\mathbf{X}, Z, Y)$.

\begin{customthm}{S1}\rm
Let $\mathcal{P}_D$ denote the law that generates i.i.d. full data random vector $D_i = (Y_i, Z_i, \mathbf{X}_i, U_i)$ and $\mathcal{H}$ the Hilbert space corresponding to all mean-zero functions of the full data with finite variance and inner product $\langle h_1, h_2 \rangle = \mathbb{E}[h_1^T(D)h_2(D)]$ where the expectation $\mathbb{E}[\cdot]$ is taken with respect to $D \sim \mathcal{P}_D$. The ortho-complement to the nuisance tangent space $\Lambda$ that corresponds to all RAL estimators in the semiparametric model $\mathcal{M}$ is 
\begin{equation*}
    \Lambda^\perp = \{h(\mathbf{X}, Z, Y) - \Pi[h(\mathbf{X}, Z, Y)\mid\Lambda], \text{where}~ h(\mathbf{X}, Z, Y) \in L^2(\mathbf{X}, Z, Y)\}.
\end{equation*}
The observed data efficient score for the semiparametric model $\mathcal{M}$ is
\begin{equation*}
    S_{\text{eff}}(\mathbf{X},Z,Y) = S_\theta(\mathbf{X}, Z, Y) - \Pi[S_\theta(\mathbf{X}, Z, Y) \mid \Lambda] = S_\theta(\mathbf{X}, Z, Y) - \mathbb{E}[a(U, \mathbf{X}) \mid \mathbf{X}, Z, Y],
\end{equation*}
where $\mathbb{E}[a(U, \mathbf{X}) \mid \mathbf{X}] = 0$, and $a(U, \mathbf{X})$ satisfies:
\begin{equation}
\label{eqn: integral eqn in H}
    \mathbb{E}[S_\theta(\mathbf{X}, Z, Y) \mid \mathbf{X}, U] = \mathbb{E}\{\mathbb{E}[a(U, \mathbf{X}) \mid \mathbf{X}, Z, Y] \mid \mathbf{X}, U\}.
\end{equation}
The semiparametric efficiency bound for $\mathcal{M}$ is hence given by $\mathbf{V}_\text{eff} = \mathbb{E}[S_{\text{eff}}(\mathbf{X},Z,Y) \cdot S^T_{\text{eff}}(\mathbf{X},Z,Y)]$. 
\label{thm: projection in H}
\end{customthm}

\begin{remark}\rm
Note that $\mathbb{E}[a(U, \mathbf{X}) \mid \mathbf{X}] = 0$ is automatically satisfied because
\begin{equation*}
    \mathbb{E}[a(U, \mathbf{X}) \mid \mathbf{X}] = \mathbb{E}\{\mathbb{E}[a(U, \mathbf{X}) \mid \mathbf{X}, Z, Y] \mid \mathbf{X}\} =
    \mathbb{E}\{S_\theta(\mathbf{X}, Z, Y) \mid \mathbf{X}\} = 0.
\end{equation*}
\end{remark}

To find the efficient score, one needs to solve the integral equation \eqref{eqn: integral eqn in H}, a task that is not operationally feasible because both $S_\theta(\mathbf{X}, Z, Y)$ and $\mathbb{E}[a(U, \mathbf{X}) \mid \mathbf{X}, Z, Y]$ depend on the unknown, ground-truth conditional distribution $f(U \mid \mathbf{X})$. To proceed, we specify a possibly incorrect \emph{working model} $f^\ast(U \mid \mathbf{X}; \xi)$ for the conditional distribution $f(U \mid \mathbf{X})$. Definition \ref{def: hilbert space star} defines a new joint law $\mathcal{P}^\ast_D$ where the ground-truth conditional law $f(U \mid \mathbf{X})$ is replaced by the possibly incorrect working model $f^\ast(U \mid \mathbf{X}; \xi)$ and the corresponding Hilbert space $\mathcal{H}^\ast$.
\begin{definition}\rm
\label{def: hilbert space star}
Let $\mathcal{H}^\ast$ denote the Hilbert space consisting of the set of all mean-zero, finite-variance $q$-dimensional vector-valued functions of the random vector $D = (Y, Z, \mathbf{X}, U) \sim \mathcal{P}_D^\ast = f(Y\mid Z, \mathbf{X}, U)\cdot f(Z \mid \mathbf{X}, U) \cdot f^\ast(U\mid \mathbf{X}; \xi) \cdot f(\mathbf{X})$. The Hilbert space $\mathcal{H}^\ast$ is endowed with the inner product $\langle h_1, h_2 \rangle = \mathbb{E}_\ast[h_1^T(D)h_2(D)]$ where the expectation $\mathbb{E}_\ast[\cdot]$ is taken with respect to the law $\mathcal{P}^\ast_D$, i.e.,
\begin{equation*}
    \mathcal{H}^\ast = \{h(D): h(D) \in \mathbb{R}^q,~ \mathbb{E}_\ast[h(D)] = 0 ~\text{and}~ \mathbb{E}_\ast[h^T(D) h(D)] < \infty \}.
\end{equation*}
\end{definition}

Although the distribution $F^\ast(U \mid \mathbf{X}; \xi)$ need not be equal to $F(U \mid \mathbf{X})$, we nevertheless require that $F(U \mid \mathbf{X})$ be absolutely continuous with respect to $F^\ast(U \mid \mathbf{X}; \xi)$, which is formally stated in Assumption \ref{assumption: absolutely continuous} below. Note that Assumption \ref{assumption: absolutely continuous} can always be satisfied by taking the support of $F^\ast(U \mid \mathbf{X}; \xi)$ to be the entire real line.

\begin{assumption}\rm
\label{assumption: absolutely continuous}
Throughout, we assume that $dF/dF^\ast < \infty$ almost surely, which essentially states that the support of our working model $f^\ast(U \mid \mathbf{X}; \xi)$ must include that of the true conditional distribution $f(U \mid \mathbf{X})$ such that for any measurable subset of $A \subseteq \mathbb{R}$, $\int_A dF^\ast(u \mid \mathbf{X}) = 0$ implies $\int_A dF(u \mid \mathbf{X}) = 0$ a.s.
\end{assumption}

Theorem \ref{thm: projection in H star} states a projection result in $\mathcal{H}^\ast$ parallel to that in Theorem \ref{thm: projection in H}.

\begin{customthm}{S2}
\label{thm: projection in H star}
\rm
Let $\mathcal{H}^\ast$ be the Hilbert space endowed with the inner product $\langle h_1, h_2 \rangle = \mathbb{E}_\ast[h_1^T(D)h_2(D)]$ as in Definition $\ref{def: hilbert space star}$, $\Lambda^\ast$ the associated nuisance tangent space and $S^\ast_\theta(\mathbf{X}, U, Z, Y)$ the corresponding full data score. The observed data efficient score is 
\begin{equation*}
    S_{\text{eff}}^\ast(\mathbf{X},Z,Y) = S^\ast_\theta(\mathbf{X}, Z, Y) - \Pi[S^\ast_\theta(\mathbf{X}, Z, Y) \mid \Lambda^\ast] = S^\ast_\theta(\mathbf{X}, Z, Y) - \mathbb{E}_\ast[a(U, \mathbf{X}) \mid \mathbf{X}, Z, Y],
\end{equation*}
where 
\begin{equation*}
    S^\ast_\theta(\mathbf{X}, Z, Y) = \mathbb{E}_\ast[S^\ast_\theta(\mathbf{X}, U, Z, Y) \mid \mathbf{X}, Z, Y],
\end{equation*}
and $a(U, \mathbf{X})$ satisfies $\mathbb{E}_\ast[a(U, \mathbf{X}) \mid \mathbf{X}] = 0$ and the integral equation:
\begin{equation}
\label{eqn: integral eqn in H star}
    \mathbb{E}_\ast[S^\ast_\theta(\mathbf{X}, Z, Y) \mid \mathbf{X}, U] = \mathbb{E}_\ast\{\mathbb{E}_\ast[a(U, \mathbf{X}) \mid \mathbf{X}, Z, Y] \mid \mathbf{X}, U\}.
\end{equation}
\end{customthm}

Unlike in Theorem \ref{thm: projection in H}, the observed data efficient score described in Theorem \ref{thm: projection in H star} is now operationally feasible as the unknown density $f(U \mid \mathbf{X})$ is replaced with the working model $f^\ast(U \mid \mathbf{X}; \xi)$ and integral equation \eqref{eqn: integral eqn in H star} can now be solved.

\begin{proof}[Theorem \ref{thm: projection in H} and \ref{thm: projection in H star}]
Recall the observed data nuisance tangent space is the projection of the full data nuisance tangent space $\Lambda^F = \Lambda_{1s}^F \oplus \Lambda_{2s}^F$ onto the observed data:
\begin{gather}
\Lambda = \Lambda_{1s} \oplus \Lambda_{2s}
\end{gather} where
\begin{gather}
\Lambda_{1} = \{a_1(\mathbf{X}): \mathbb{E}[a_1(\mathbf{X})] = 0\}\\
\Lambda_{2} = \{\mathbb{E}[a_2(U, \mathbf{X}) \mid \mathbf{X}, Z, Y]: \mathbb{E}[a_2(U, \mathbf{X}) \mid \mathbf{X}] = 0\}.
\end{gather}
We project the observed-data score $S_\theta(\mathbf{X}, Z, Y)
= ~\mathbb{E}[S_\theta(\mathbf{X}, U, Z, Y) \mid \mathbf{X}, Z, Y]$ onto $\Lambda$. Note the observed-data score satisfies $\mathbb{E}[S_\theta(\mathbf{X}, Z, Y) \mid \mathbf{X}] = 0$, and $\mathbb{E}[S_\theta(\mathbf{X}, Z, Y)] \perp \Lambda_{1}$. Elements $h(\mathbf{X},Z,Y)$ orthogonal to $\Lambda_{2}$ satisfy \[
\mathbb{E}[h^T(\mathbf{X}, Z, Y) \cdot \mathbb{E}[a_2(U, \mathbf{X}) \mid \mathbf{X}, Z, Y]] = 0
\] for all $a_2(U, \mathbf{X})$ such that $\mathbb{E}[a_2(U, \mathbf{X}) \mid \mathbf{X}] = 0$. By iterated expectation, we have 
\begin{equation}
\begin{split}
0 &= \mathbb{E}[h^T(\mathbf{X}, Z, Y) \cdot \mathbb{E}[a_2(U, \mathbf{X}) \mid \mathbf{X}, Z, Y]] \\
&= \mathbb{E}[\mathbb{E}[h^T(\mathbf{X}, Z, Y)\cdot a_2(U, \mathbf{X}) \mid \mathbf{X}, Z, Y]] \\
&= \mathbb{E}[h^T(\mathbf{X}, Z, Y)\cdot a_2(U, \mathbf{X})]
\end{split}
\end{equation}
Hence, any element $h(\mathbf{X},Z,Y)$ satisfying $\mathbb{E}[h^T(\mathbf{X}, Z, Y) \mid U, \mathbf{X}] = 0$ is orthogonal to $\Lambda_{2}$. Finally, the projection of $S_\theta(\mathbf{X}, Z, Y)$ onto $\Lambda^\perp_{2}$ is $S_\theta(\mathbf{X}, Z, Y) - \mathbb{E}[a_2(U, \mathbf{X}) \mid \mathbf{X}, Z, Y]$ with a properly chosen $a_2(U, \mathbf{X})$ such that $\mathbb{E}[a_2(U, \mathbf{X}) \mid \mathbf{X}] = 0$, and satisfies $\mathbb{E}\{S_\theta(\mathbf{X}, Z, Y) - \mathbb{E}[a_2(U, \mathbf{X}) \mid \mathbf{X}, Z, Y] \mid U, \mathbf{X}\} = 0$, or equivalently: \[
\mathbb{E}[S_\theta(\mathbf{X}, Z, Y) \mid U, \mathbf{X}] = \mathbb{E}\{\mathbb{E}[a_2(U, \mathbf{X}) \mid \mathbf{X}, Z, Y] \mid U, \mathbf{X}\}.
\] 
This concludes the proof of Theorem \ref{thm: projection in H}. Proving Theorem \ref{thm: projection in H star} is identical, except that we need to change the Hilbert space from $\mathcal{H}$ to $\mathcal{H}^\ast$, and replace $\mathbb{E}[\cdot]$ with $\mathbb{E}_\ast[\cdot]$.
\end{proof}

\bigskip
\bigskip
\bigskip
\begin{center}
{\large\bf Supplementary Material B: Numerical Solution of the Integral Equation and Computation}
\end{center}
\bigskip

To calculate the observed data efficient score according to Theorem \ref{thm: projection in H star}, we need to find $a(U, \mathbf{X})$ at each observed value $\mathbf{X} = \mathbf{X}_i$ such that the integral equation \eqref{eqn: integral eqn in H star} holds. Theorem \ref{thm: fredholm equivalence} shows that solving equation \eqref{eqn: integral eqn in H star} is equivalent to solving a Fredholm integral equation of the first kind.

\begin{customthm}{S3}\rm
\label{thm: fredholm equivalence}
Let $\Omega(\cdot)$ denote the domain of the corresponding variable. Solving Equation (\ref{eqn: integral eqn in H star}) is equivalent to solving the following Fredholm integral equation of the first kind over the domain of $u$:
\begin{equation}
    \int_{\Omega(u')} a(u', \mathbf{X}) \cdot K(u', u, \mathbf{X})~d\mu(u') = b(u, \mathbf{X}),
\label{eqn: Fredholm integral equation}
\end{equation}
with the forcing function $b(u, \mathbf{X})$ and kernel $K(u', u, \mathbf{X})$ defined as follows:
\begin{equation*}
\begin{split}
    &b(u, \mathbf{X}) = \int_{\Omega(y) \times \Omega(z)} \frac{\int S_\beta(y, z, \mathbf{X}, u') f(y,z \mid \mathbf{X}, u') f^\ast(u' \mid \mathbf{X}; \xi) ~d\mu(u')}{I^\ast(y, z, \mathbf{X})} \\
    & \hspace{7cm}\times f(y, z\mid \mathbf{X}, U = u)~d(\nu_1 \times \nu_2)(y, z), \\
    &K(u', u, \mathbf{X}) = f^\ast(u' \mid \mathbf{X}; \xi) \int_{\Omega(y) \times \Omega(z)} \frac{ f(y, z\mid \mathbf{X}, u')\cdot f(y, z\mid \mathbf{X}, u)}{I^\ast(y, z, \mathbf{X})}~d(\nu_1 \times \nu_2)(y, z),\\
    &I^\ast(y, z, \mathbf{X}) = \int_{\Omega(u')} f(y, z \mid \mathbf{X}, u')\cdot f^\ast(u' \mid \mathbf{X}; \xi)~d\mu(u').
\end{split}
\end{equation*}
\end{customthm}

\begin{proof}[Theorem \ref{thm: fredholm equivalence}]
Fix a data point $(Y, Z, \mathbf{X})$, and $U = u$, we write 
\[
\mathbb{E}[S^\ast_\theta(Y, Z, \mathbf{X}) \mid \mathbf{X}, U = u] = \int S^\ast_\theta(y, z, \mathbf{X}) \cdot f(y, z \mid \mathbf{X}, U = u; \lambda, \beta, \kappa, c_\delta, c_\gamma)~d\nu(y)~d\nu(z),
\] and
\begin{equation*}
    \begin{split}
        &\mathbb{E}\{\mathbb{E}_\ast[a(\mathbf{X}, U) \mid \mathbf{X}, Z, Y] \mid \mathbf{X}, U = u\} \\
        &= \int \mathbb{E}_\ast[a(\mathbf{X}, U) \mid \mathbf{X}, z, y] \cdot f(y, z \mid \mathbf{X}, U = u; \lambda, \beta, \kappa, c_\delta, c_\gamma)~d\nu(y)~\nu(z),
    \end{split}
\end{equation*}
where $f(y, z \mid \mathbf{X}, U = u)$ is the conditional distribution of $(Y = y, Z = z)$ given $\mathbf{X}$ and $U = u$.

We compute:
\begin{equation*}
    \begin{split}
        S_\theta^\ast(Y, Z ,\mathbf{X}) &= \mathbb{E}_\ast\{S_\theta(Y, Z, \mathbf{X}, U) \mid Y, Z, \mathbf{X}\} \\ 
        &= \frac{\int S_\theta(Y, Z, \mathbf{X}, u) f(Y, Z \mid \mathbf{X}, u; \lambda, \beta, \kappa, c_\delta, c_\gamma)f^\ast(u \mid \mathbf{X}; \xi)f(\mathbf{X})~d\mu(u)}{\int f(Y, Z \mid \mathbf{X}, u;\lambda, \beta, \kappa, c_\delta, c_\gamma) f^\ast(u \mid \mathbf{X}; \xi) f(\mathbf{X}) ~d\mu(u)}\\
        &= \frac{\int S_\theta(Y, Z, \mathbf{X}, u) f(Y, Z \mid \mathbf{X}, u; \lambda, \beta, \kappa, c_\delta, c_\gamma)f^\ast(u \mid \mathbf{X}; \xi)~d\mu(u)}{\int f(Y, Z \mid \mathbf{X}, u;\lambda, \beta, \kappa, c_\delta, c_\gamma) f^\ast(u \mid \mathbf{X}; \xi) ~d\mu(u)},
    \end{split}
\end{equation*}
where $f^\ast(U \mid \mathbf{X}; \xi)$ is a user-supplied, possibly incorrectly, conditional density.

On the other hand, we have:
\begin{equation*}
    \begin{split}
        \mathbb{E}_\ast[a(\mathbf{X}, U) \mid \mathbf{X}, Z, Y] = \frac{\int a(\mathbf{X}, U)\cdot f(Y, Z \mid \mathbf{X}, u; \lambda, \beta, \kappa, c_\delta, c_\gamma)\cdot f^\ast(u \mid \mathbf{X}; \xi)~d\mu(u)}{\int f(Y, Z \mid \mathbf{X}, u; \lambda, \beta, \kappa, c_\delta, c_\gamma)\cdot f^\ast(u \mid \mathbf{X}; \xi)~d\mu(u)}.
    \end{split}
    \label{eqn: E_ast(a_2(Z,X,U|X,Z,Y))}
\end{equation*}

Put together, we have
\begin{equation*}
\begin{split}
    &\mathbb{E}[S^\ast_\theta(Y, Z, \mathbf{X}) \mid \mathbf{X}, U = u] \\
    &= \int S^\ast_\theta(y, z, \mathbf{X}) \cdot f(y, z \mid \mathbf{X}, U = u)~d\nu(y)~d\nu(z) \\
    &= \int \frac{\int S_\theta(y, z, \mathbf{X}, u') f(y,z \mid \mathbf{X}, u'; \lambda, \beta, \kappa, c_\delta, c_\gamma) f^\ast(u' \mid \mathbf{\mathbf{X}}; \xi) ~d\mu(u')}{\int f(y, z \mid \mathbf{X}, u''; \lambda, \beta, \kappa, c_\delta, c_\gamma) f^\ast(u'' \mid \mathbf{X}; \xi) ~d\mu(u'')} \\
    &\hspace{2cm}\times f(y, z \mid \mathbf{X}, U = u; \lambda, \beta, \kappa, c_\delta, c_\gamma)~d\nu(y)~d\nu(z) \\
    &= b(u, \mathbf{\mathbf{X}}),
\end{split}
\end{equation*}
and
\begin{equation*}
    \begin{split}
        &\mathbb{E}\{\mathbb{E}_\ast[a(\mathbf{X}, U) \mid \mathbf{X}, Z, Y] \mid \mathbf{X}, U = u\} \\
        &= \int \mathbb{E}_\ast[a(\mathbf{X}, U) \mid \mathbf{X}, z, y] \cdot f(y, z \mid \mathbf{X}, U = u; \lambda, \beta, \kappa, c_\delta, c_\gamma)~d\nu(y)~d\nu(z) \\
        &=\int \frac{\int a(\mathbf{X}, u')\cdot f(y, z \mid \mathbf{X}, u'; \lambda, \beta, \kappa, c_\delta, c_\gamma)\cdot f^\ast(u' \mid \mathbf{\mathbf{X}}; \xi)~d\mu(u')}{\int f(y, z \mid \mathbf{X}, u''; \lambda, \beta, \kappa, c_\delta, c_\gamma)\cdot f^\ast(u'' \mid \mathbf{X}; \xi)~d\mu(u'')} \\
        &\hspace{5cm}\times f(y, z \mid \mathbf{X}, U = u; \lambda, \beta, \kappa, c_\delta, c_\gamma)~d\nu(y)~d\nu(z) \\
         &=\int \frac{\int a(\mathbf{X}, u')\cdot f(y, z \mid \mathbf{X}, u'; \lambda, \beta, \kappa, c_\delta, c_\gamma)\cdot f^\ast(u' \mid \mathbf{\mathbf{X}}; \xi)~d\mu(u')}{I^\ast(y, z, \mathbf{\mathbf{X}})} \\
         &\hspace{5cm}\times f(y, z \mid \mathbf{X}, U = u; \lambda, \beta, \kappa, c_\delta, c_\gamma)~d\nu(y)~d\nu(z) \\
        &=\int a(\mathbf{X}, u')\underbrace{f^\ast(u' \mid \mathbf{\mathbf{X}}; \xi) \int\frac{ f(y, z \mid \mathbf{X}, u'; \lambda, \beta, \kappa, c_\delta, c_\gamma)\cdot f(y, z \mid \mathbf{X}, u; \lambda, \beta, \kappa, c_\delta, c_\gamma)}{I^\ast(y, z, \mathbf{\mathbf{X}})}d\nu(y)d\nu(z)}_{K(u', u, \mathbf{\mathbf{X}})}d\mu(u'),
    \end{split}
\end{equation*}
where we denote $I^\ast(y, z, \mathbf{X}) = \int f(y, z \mid \mathbf{X}, u; \lambda, \beta, \kappa, c_\delta, c_\gamma)\cdot f^\ast(u \mid \mathbf{X}; \xi)~d\mu(u)$, and assume the integrand is absolutely integrable with respect to the product measure $\mu(u') \times \nu(y) \times \nu(z)$ and apply the Fubini's theorem.

Put together, solving for $a(\mathbf{X}, U)$ is equivalent to solving the following Fredholm's integral equation of the first kind over the domain of $u$, $\Omega(u) = [0, 1]$:
\begin{equation*}
    \int_{\Omega(u')} a(u', \mathbf{X}) \cdot K(u', u, \mathbf{X})~d\mu(u') = b(u, \mathbf{X}),
\end{equation*}
where the kernel $K(u', u, \mathbf{X})$ and the forcing function $b(u, \mathbf{X})$ are defined above, and the equation is solved for each data point $\mathbf{X}_i$ in the dataset.
\end{proof}

We now describe formal conditions for the existence of a solution to integral equation \eqref{eqn: Fredholm integral equation}. We restrict our attention to cases when the integral operator induced by $K(u', u, \mathbf{X})$ is compact. Note when both $Y$ and $Z$ are binary, the kernel becomes degenerate:
\begin{equation*}
    K(u', u, \mathbf{X}) = \sum_{i,j} g_{ij}(\mathbf{X}, u') \times h_{ij}(\mathbf{X}, u),
\end{equation*}
where $g_{ij}(\mathbf{X}, u') = f^\ast(u' \mid \mathbf{X};\xi) \cdot f(y = i, z = j \mid \mathbf{X}, u')/I^\ast(y = i, z = j, \mathbf{X})$ and $h_{ij}(\mathbf{X}, u) = f(y = i, z = j \mid \mathbf{X}, u)$ for $(i, j) \in \{0, 1\} \times \{0, 1\}$. Therefore, this degenerate kernel is bounded and the induced integral operator as long as $g_{ij} \in L^2(u', \mathbf{X})$ and $h_{ij} \in L^2(u, \mathbf{X})$ (\citealp{carrasco2007linear}). When the kernel is not degenerate, a sufficient condition for the operator to be compact is that $K$ is a Hilbert-Schmidt kernel, i.e., $K \in L^2(u', u, \mathbf{X})$, i.e.,
\begin{equation*}
\begin{split}
    &\int_{\Omega(u'), \Omega(u)} f^\ast(u' \mid \mathbf{X}; \xi)^2 \left(\int_{\Omega(y) \times \Omega(z)} \frac{ f(y, z \mid \mathbf{X}, u')\cdot f(y, z \mid \mathbf{X}, u)}{I^\ast(y, z, \mathbf{X})}~d(\nu_1 \times \nu_2)(y, z)\right)^2\\
    &\hspace{10cm} \cdot d(\mu_1 \times \mu_2)(u', u) < \infty.
\end{split}
\end{equation*}

For the compact integral operator $K$, there exists a singular system $(\lambda_n, \phi_n, \psi_n),~n = 1,...,\infty$ of $K$ with nonzero singular values $\{\lambda_n\}$ and orthogonal sequences $\{\phi_n\}$ and $\{\psi_n\}$ such that $K\phi_n = \lambda_n \psi_n$ and $K^\ast \psi_n = \lambda_n \phi_n$ (\citealp{kress1989linear}). Moreover, the Picard's theorem (\citealp{kress1989linear}) states that the integral equation (\ref{eqn: Fredholm integral equation}) $Ka = b$ is solvable if and only if 
\begin{enumerate}
    \item $b \in \mathcal{N}(K^\ast)^\perp$,
    \item $\sum_{n = 1}^\infty \lambda_n^{-2} \|\langle b, \psi_n \rangle\|^2 < \infty$.
\end{enumerate}
Note when $u'$ and $u$ are discrete and the kernel reduces to a matrix, above conditions reduces to the kernel matrix has full rank and is invertible.


A standard approach to solve the Fredholm equation \eqref{eqn: Fredholm integral equation} is to express the integral in terms of a Gauss-type quadrature formula and obtain approximate ``pivotal'' values of $a(u', \mathbf{X})$, i.e., $a(u'_0, \mathbf{X}), a(u'_2, \mathbf{X}), ..., a(u'_K, \mathbf{X})$, from a set of linear simultaneous equations (\citet{baker1964numerical}). Let $\mathcal{P}_U = (u_0, u_1, ..., u_K)$ such that $0 = u_0 < u_1 < u_2 < ... < u_K = 1$ be an equal-spaced partition of $\Omega(u)$. For a fixed $u_j$ and a partition $\mathcal{P}_{U'} = (u'_0, u'_1, ..., u'_K)$ such that $0 = u'_0 < u'_1 < ... < u'_K = 1$ of $\Omega(u')$, we can approximate the integral in \eqref{eqn: Fredholm integral equation} by:
\begin{equation}
\begin{split}
    b(u_j, \mathbf{X}) &= \int_{\Omega(u')} a(u', \mathbf{X}) \cdot K(u', u, \mathbf{X})~d\mu(u') \\
    &= \sum_{i = 0}^K h\cdot w_i \cdot a(u'_i, \mathbf{X}) \cdot K(u'_i, u_j, \mathbf{X}) + e,
    \label{eqn: newton-cortes formula}
\end{split}
\end{equation}
where $h$ is the mesh size of $\mathcal{P}_{U'}$, $w_i$ are weights in the Newton-Cotes formula, and $e$ captures the approximation error. Let $\mathbf{K}$ be a $(K+1) \times (K+1)$ matrix with $ij$-th entry $K(u'_i, u_j, \mathbf{X})$, $\mathbf{W} = \text{diag}\{w_0, w_1, ..., w_K\}$, $\mathbf{a}$ and $\mathbf{b}$ both $(K+1) \times 1$ vector with $i$-th element $b(u_i, \mathbf{X})$ and $a(u'_i, \mathbf{X})$. In matrix notation, we can write Equation \eqref{eqn: newton-cortes formula} as 
\begin{equation*}
    h\mathbf{K}^T \mathbf{W} \mathbf{a} + e = \mathbf{b},
\end{equation*}
where $\mathbf{a}$ is to be solved. Note that $\mathbf{W} = \text{diag}\{1/2, 1, ..., 1, 1/2\}$ corresponds to the trapezoid rule.

It is well-understood that Fredholm equation of the first kind, despite admitting a unique solution, can be ill-posed and unstable, and the associated system of linear simultaneous equations can be ill-posed as well (\citealp{phillips1962technique,baker1964numerical}). To overcome this difficulty, Equation \eqref{eqn: Fredholm integral equation} may be transformed into an approximation Fredholm equation:
\begin{equation}
    b(u, \mathbf{X}) - \int_{\Omega(u')} a_\alpha(u', \mathbf{X}) \cdot K(u', u, \mathbf{X})~d\mu(u') = \alpha a_\alpha(u, \mathbf{X}),
    \label{eqn: Fredholm of second kind}
\end{equation}
which is a well-posed Fredholm equation of the second kind. In Equation \eqref{eqn: Fredholm of second kind}, $\alpha$ is a small positive \emph{regularization parameter}. It has been shown that 
\[
\lim_{\alpha \rightarrow 0}a_\alpha(u, \mathbf{X}) = a(u, \mathbf{X})
\] by \citet{tikhonov1963solution} and \citet{phillips1962technique}. We will approximate $a(u, \mathbf{X})$ by solving Equation \eqref{eqn: Fredholm of second kind} for some small $\alpha$ value, which is equivalent to minimizing a Ridge-regression-type of loss with regularization parameter $\alpha$, i.e., $\|h\mathbf{K}^T \mathbf{W} \mathbf{a} - \mathbf{b} \|_2^2 + \alpha \|\mathbf{a}\|_2^2$.

\bigskip
\bigskip
\bigskip
\begin{center}
{\large\bf Supplementary Material C: Proofs, Derivations, and Conditions}
\end{center}

\subsection*{C.1: Proof of Proposition 1}
When $Y$, $Z$, and $U$ are all binary and there is no $U, Z$ interaction, their joint distribution can be parametrized in the following way:
\begin{equation*}
\begin{split}
    f(Y, Z, U) ~\propto~ &f(Y \mid Z = 0, U = 0)\cdot f(Z \mid Y = 0, U = 0) \cdot f(U \mid A = 0, Y = 0)\cdot\\ 
    &\text{OR}(Y, Z \mid U = 0)\cdot \text{OR}(Y, U \mid Z = 0)\cdot \text{OR}(Z, U \mid Y = 0).
    \end{split}
\end{equation*}
Note \[
\exp(\beta_0 Y) = \frac{f(Y \mid Z = 0, U = 0)}{f(Y = 0 \mid Z = 0, U = 0)}; 
~~\exp(\alpha_0 Z) = \frac{f(Z \mid Y = 0, U = 0)}{f(Z = 0 \mid Y = 0, U = 0)};
\] and we have
\[
f(Y, Z, U) \propto \exp\{\beta_0 Y + \alpha_0 Z + \beta_z Y Z + c_\delta Y U + c_\gamma U Z\}.
\]
Therefore, we have 
\begin{equation*}
    \begin{split}
        L(Y, Z) &= \frac{1}{C} \int \exp\{\beta_0 Y + \alpha_0 Z + \beta_z Y Z + c_\delta Y u + c_\gamma u Z\} f(u \mid Z = 0, Y = 0)du \\
        &=\frac{1}{C} \exp\{\beta_0 Y + \alpha_0 Z + \beta_z Y Z\}\cdot \int \exp\{c_\delta Y u + c_\gamma u Z\} f(u \mid Z = 0, Y = 0)du \\
        &=\frac{1}{C}  \exp\{\beta_0 Y + \alpha_0 Z + \beta_z Y Z\}\cdot \mathcal{M}_{U \mid Z = 0, Y = 0}(c_\delta Y + c_\gamma Z),
    \end{split}
\end{equation*}
where $\mathcal{M}_{U \mid Z = 0, Y = 0}(c_\delta Y + c_\gamma Z)$ is the moment generating function of $f(U \mid Y, Z)$ evaluated at $c_\delta Y + c_\gamma Z$ and $(c_\delta, c_\gamma)$ are fixed constants.

The observed data are $L(Y = 0, Z = 0)$, $L(Y = 0, Z = 1)$, $L(Y = 1, Z = 0)$, and $L(Y = 1, Z = 1)$, which implies:
\begin{align*}
    &L(0, 0) = \frac{1}{C} \mathcal{M}_{U \mid Z = 0, Y = 0}(0) ~\Rightarrow~ C = 1/L(0, 0), \\
    &L(0, 1) = \frac{1}{C} \exp(\alpha_0) \cdot \mathcal{M}_{U\mid Z = 0, Y = 0}(c_\gamma) \Rightarrow \alpha_0 = \log\left(\frac{L(0, 1)}{L(0, 0)\mathcal{M}_{U \mid Z = 0, Y = 0}(c_\gamma)} \right),\\
    &L(1, 0) = \frac{1}{C} \exp(\beta_0) \cdot \mathcal{M}_{U \mid Z = 0, Y = 0}(c_\delta) \Rightarrow \beta_0 = \log\left(\frac{L(1, 0)}{L(0, 0)\mathcal{M}_{U \mid Z = 0, Y = 0}(c_\delta)} \right), \\
    &L(1, 1) = \frac{1}{C} \exp(\beta_0 + \alpha_0 + \beta_z) \cdot \mathcal{M}_{U \mid Z = 0, Y = 0}(c_\delta + c_\gamma) \\
    &\Rightarrow \beta_z = \log\left(
    \frac{L(1, 1)}{L(0, 1)L(1, 0)} \cdot \frac{\mathcal{M}_{U\mid Z = 0, Y = 0}(c_\delta)\mathcal{M}_{U \mid Z = 0, Y = 0}(c_\gamma)}{\mathcal{M}_{U \mid Z = 0, Y = 0}(c_\delta + c_\gamma)}\right).
\end{align*}

Therefore, we see observed data plus the underlying distribution of $U$ uniquely identify parameters $\beta_0$, $\alpha_0$, and $\beta_z$.

\subsection*{C.2: Proof of Proposition 2}
$S^\ast_{\text{eff}}(\mathbf{X}, Z, Y)$ is the efficient score constructed according to Theorem \ref{thm: projection in H star}; therefore, it necessarily satisfies 
\[
 \mathbb{E}_\ast[S^\ast_{\text{eff}}(\mathbf{X},Z,Y) \mid \mathbf{X}, U] = 0.
\]

The conditional distribution of $(\mathbf{X}, Z, Y)$ given $(\mathbf{X}, U)$ depends on $Z \mid \mathbf{X}, U$ and $Y \mid \mathbf{X}, U, Z$, both of which are assumed to be correctly specified. Therefore, $\mathbb{E}_\ast[h^T(\mathbf{X}, Z, Y) \mid U, \mathbf{X}] = 0$ implies $\mathbb{E}[h^T(\mathbf{X}, Z, Y) \mid U, \mathbf{X}] = 0$ for any random function $h(\mathbf{X}, Z, Y)$. Apply this result to $S^\ast_{\text{eff}}(\mathbf{X},Z,Y) \mid \mathbf{X}, U$ and we see immediately 
\[
 \mathbb{E}[S^\ast_{\text{eff}}(\mathbf{X},Z,Y) \mid \mathbf{X}, U] = 0.
\]
$\mathbb{E}[S^\ast_{\text{eff}}(\mathbf{X},Z,Y)] = 0$ then follows.

\subsection*{C.3: Proof of Theorem 1}
The consistency and asymptotic normality of the estimator follows from standard semiparametric theory. To prove the consistency, it suffices to show $\mathbb{E}\{S^\ast_{\text{eff}}(\mathbf{X},Z,Y; \theta_0)\} = 0$, together with the following regularity conditions on the smoothness of the Jacobian and its limit (\citealp{Foutz1977}):
\begin{enumerate}
    \item $n^{-1}\sum_{i = 1}^n \partial S^\ast_{\text{eff}}(\mathbf{X}_i, Z_i, Y_i; \theta)/\partial \theta$ exists and is continuous in an open neighborhood of $\beta_0$;
    \item $n^{-1}\sum_{i = 1}^n \partial S^\ast_{\text{eff}}(\mathbf{X}_i, Z_i, Y_i; \theta)/\partial \theta$ converges uniformly to its limit in a neighborhood of $\theta_0$;
    \item $\mathbb{E}\{ \partial S^\ast_{\text{eff}}(\mathbf{X}, Z, Y; \theta)/\partial \theta\}|_{\theta = \theta_0}$ is invertible.
\end{enumerate}

Denote $J_n(\theta) = n^{-1}\sum_{i = 1}^n \partial S^\ast_{\text{eff}}(\mathbf{X}_i, Z_i, Y_i; \theta)/\partial \theta$. Let the solution to the estimating equation be $\hat{\theta}_n$ and $\Tilde{\theta}_n$ a value between $\theta_0$ and $\hat{\theta}_n$. We have 
\begin{equation}
    \begin{split}
        n^{1/2}(\hat{\theta}_n - \theta_0) &= -J_n(\Tilde{\theta}_n)^{-1} \frac{1}{\sqrt{n}} \sum_{i = 1}^n S^\ast_{\text{eff}}(\mathbf{X}_i, Z_i, Y_i; \theta_0) \\
        &= -\mathbb{E}\{\partial S^\ast_{\text{eff}}(\mathbf{X}_i, Z_i, Y_i; \theta_0)/\partial \theta\}^{-1}\frac{1}{\sqrt{n}} \sum_{i = 1}^n S^\ast_{\text{eff}}(\mathbf{X}_i, Z_i, Y_i; \theta_0) + o_p(1).
    \end{split}
\end{equation}

Note $\hat{\theta}_n$ is asymptotically normal with variance covariance matrix \[
\mathbf{V} = \mathbb{E}\{\partial S^\ast_{\text{eff}}(O_i; \theta_0)/\partial \theta\}^{-1} \mathbb{E}\{S^\ast_{\text{eff}}(O_i; \theta_0)S^\ast_{\text{eff}}(O_i; \theta_0)^T\}\mathbb{E}\{\partial S^\ast_{\text{eff}}(O_i; \theta_0)/\partial \theta^T\}^{-1},
\]
where $O_i = (\mathbf{X}_i, Z_i, Y_i)$.

\bigskip
\bigskip
\bigskip
\begin{center}
{\large\bf Supplementary Material D: Uniformly Valid Confidence Band}
\end{center}

\subsection*{D.1: Constructing a Uniformly Valid Confidence Band of $\beta_{c_\gamma, c_\delta}$}
In this section, we show how one can easily construct a confidence band for $\beta(c_\delta, c_\gamma)$, viewed as a function in $(c_\delta, c_\gamma)$, that is uniformly valid for $(c_\delta, c_\gamma) \in \Delta \times \Gamma$ leveraging a version of the multiplier bootstrap. Below, we will use $\hat{\beta}(c_\delta, c_\gamma)$ to denote an estimate for $\beta(c_\delta, c_\gamma)$.

\begin{customprop}{S1}\rm
\label{prop: confidence band}
Let $\hat{\theta}(c_\delta, c_\gamma)$ be constructed as in Theorem 1 and $\hat{\beta}(c_\delta, c_\gamma)$ the coordinate of $\hat{\theta}(c_\delta, c_\gamma)$ that corresponds to $\beta$. Let $V(c_\delta, c_\gamma)$ be the variance of $\beta(c_\delta, c_\gamma)$, and $\hat{V}(c_\delta, c_\gamma)$ a consistent estimator of $\mathbf{V}(c_\delta, c_\gamma)$. Let $\epsilon_i, i = 1,2, ..., N$ be independent $\text{Normal}(0, 1)$ random variables independent of the data. Define 
\begin{equation*}
  Z_{\infty}^{\text{MB}} = \sup_{(c_\delta, c_\gamma) \in \Delta \times \Gamma} \left| \frac{1}{\sqrt{n}} \sum_{i = 1}^N \frac{\epsilon_i \cdot \hat{S}^\ast_{\text{eff}, 1}(\mathbf{X}_i, Z_i, Y_i) }{\sqrt{\hat{V}(c_\delta, c_\gamma)}} \right|,  
\end{equation*}
where $\hat{S}^\ast_{\text{eff}, 1}(\mathbf{X}_i, Z_i, Y_i)$ denotes the coordinate of $\hat{S}^\ast_{\text{eff}}(\mathbf{X}_i, Z_i, Y_i) = S^\ast_{\text{eff}}(\mathbf{X}_i, Z_i, Y_i; \hat{\theta})$ corresponding to $\hat{\beta}$. Let $\hat{c}_{1 - \alpha}$ be the conditional $1 - \alpha$ quantile of $Z_{\infty}^{\text{MB}}$ given the data. Under certain regularity conditions,
\[
P\left(\beta(c_\delta, c_\gamma) \in \left[\hat{\beta}(c_\delta, c_\gamma) - \hat{c}_{1 - \alpha} \sqrt{\hat{V}(c_\delta, c_\gamma)}, ~\hat{\beta}(c_\delta, c_\gamma) + \hat{c}_{1 - \alpha} \sqrt{\hat{V}(c_\delta, c_\gamma) }\right]~\forall~ (c_\delta, c_\gamma) \in \Delta \times \Gamma \right)
\] 
converges to $1 - \alpha$ as $N \rightarrow \infty$.
\end{customprop}

Proposition \ref{prop: confidence band} is a straightforward application of the multiplier bootstrap technique. For more details on regularity conditions, see \citet{vaart1996weak}.

\subsection*{D.2: War and Political Participation Example: a Uniformly Valid Confidence Band}
\begin{figure}%
  \centering
  \subfigure[\small Distribution of $Z_\infty^{\text{TB}}$]{%
     \label{fig: dist of z_tb}%
    \includegraphics[scale = 0.38]{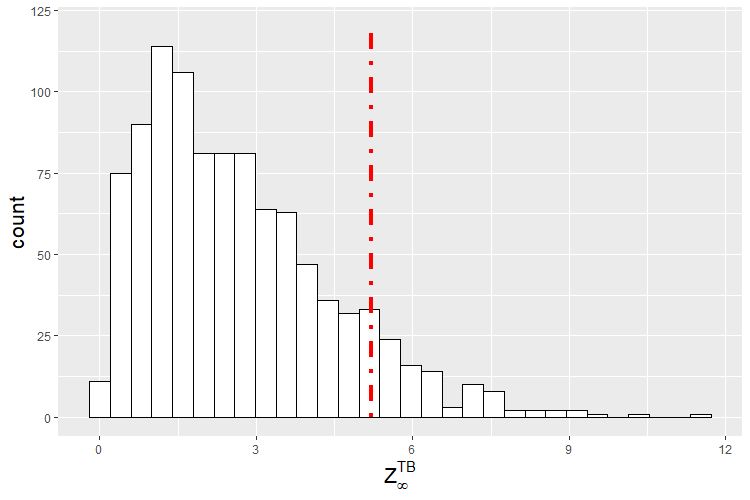}}%
  \subfigure[\small Confidence band of $\hat{\beta}_{c_\delta, c_\gamma}$]{%
    \label{fig: conf band}%
    \includegraphics[scale = 0.38]{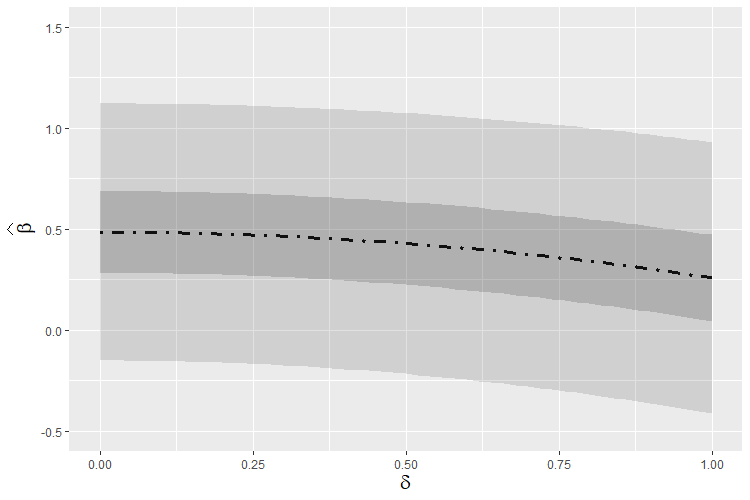}}
    \caption{\small A uniformly valid confidence band for a binary $U$ and $c_\gamma = c_\delta \in [0, 1.0]$. Left panel is the distribution of multiplier bootstrapped statistic $Z_\infty^{\text{TB}}$, with the red vertical line being $0.90$ quantile. Right panel is the estimate with uncertainty quantification. The light shade is a level $0.1$ uniformly valid confidence band; the darker shade is pointwise level $0.1$ confidence intervals.}
  \label{fig: binary U conf band}
\end{figure}

Sometimes, practitioners would like to specify a range of $(c_\delta, c_\gamma)$, say $c_\delta = c_\gamma \in [0, \Tilde{\delta}]$, where we let $c_\delta = c_\gamma$ for simplicity, and construct a confidence band that takes into account this uncertainty of sensitivity parameters. As discussed in Section 7, a straightforward application of the multiplier bootstrap technique serves this purpose. We demonstrate how it works using the war and political participation data. For the purpose of illustration, we only perform one single imputation and illustrate the method using this imputed dataset. For a binary $U$ and $\Tilde{\delta} = 1.0$, left panel of Figure \ref{fig: conf band} displays the distribution of the multiplier bootstrapped statistic $Z_\infty^{\text{TB}}$ using $1000$ resamples, and right panel displays a level $0.1$ uniformly valid confidence band for $\hat{\beta}_{c_\delta, c_\gamma}$ on the line $c_\delta = c_\gamma$ for $c_\delta \in [0, 1.0]$. To draw a contrast, we also impose a pointwise confidence interval on the same plot (the darker shade). Note a uniformly valid confidence band is significantly more conservative.

\clearpage
\begin{center}
{\large\bf Supplementary Materials E: Additional Simulation Results}
\end{center}

\subsection*{E.1: Binary U and Binary Y: Plots of Monte Carlo Distributions hen n = 300 and n = 500}

\begin{figure}[h]
  \centering
  \subfigure[\small$\hat{\beta}^\ast_{\text{semi}}$: incorrectly specified $U$, $n = 300$]{%
    \label{app fig: binary U binary Y incorrect U n = 300}%
    \includegraphics[height = 5 cm, width = 0.5\linewidth]{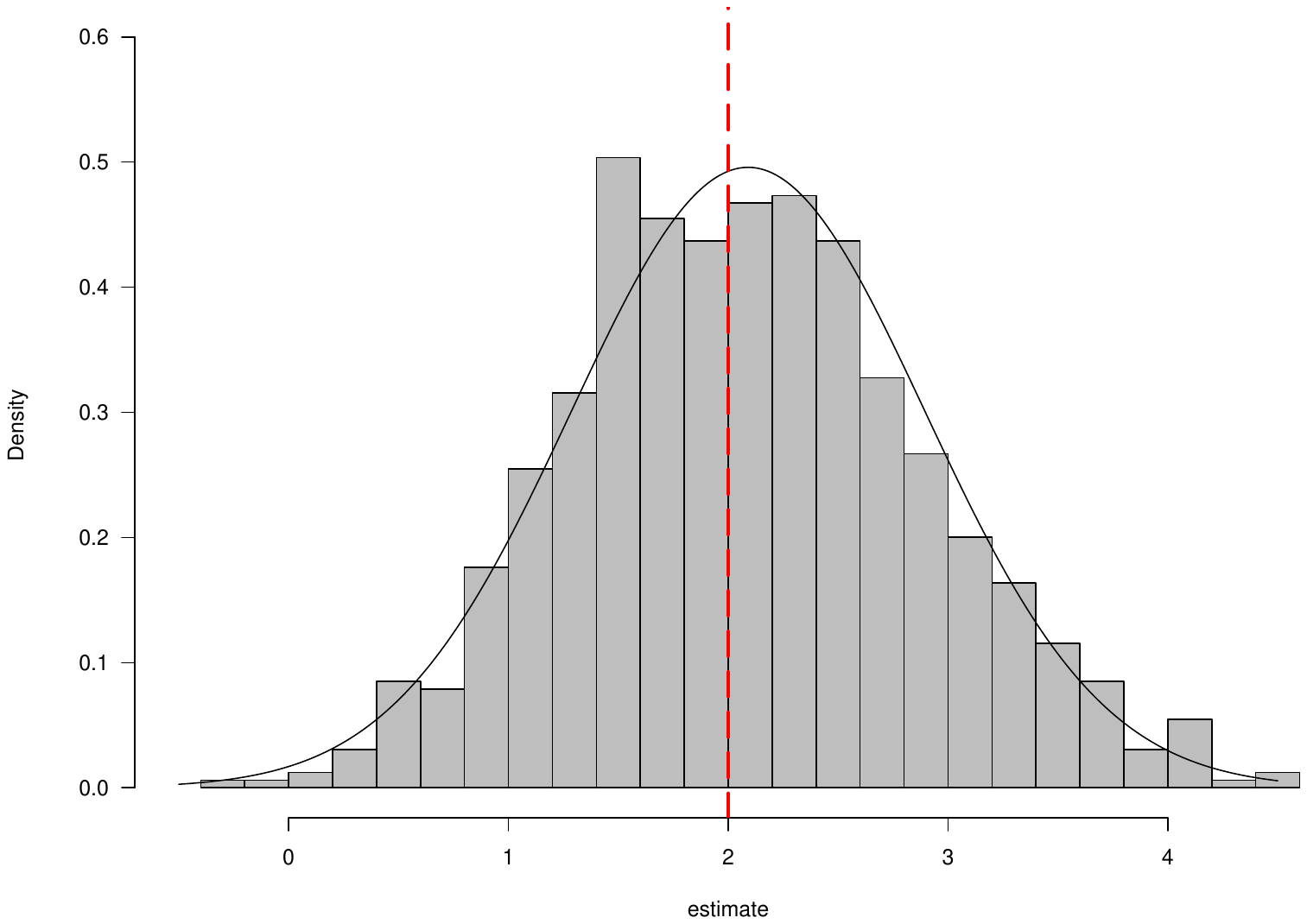}}%
  \subfigure[\small$\hat{\beta}_{\text{semi}}$: correctly specified $U$, $n = 300$]{%
    \label{app fig: binary U binary Y correct U n = 300}%
    \includegraphics[height = 5 cm, width = 0.5\linewidth]{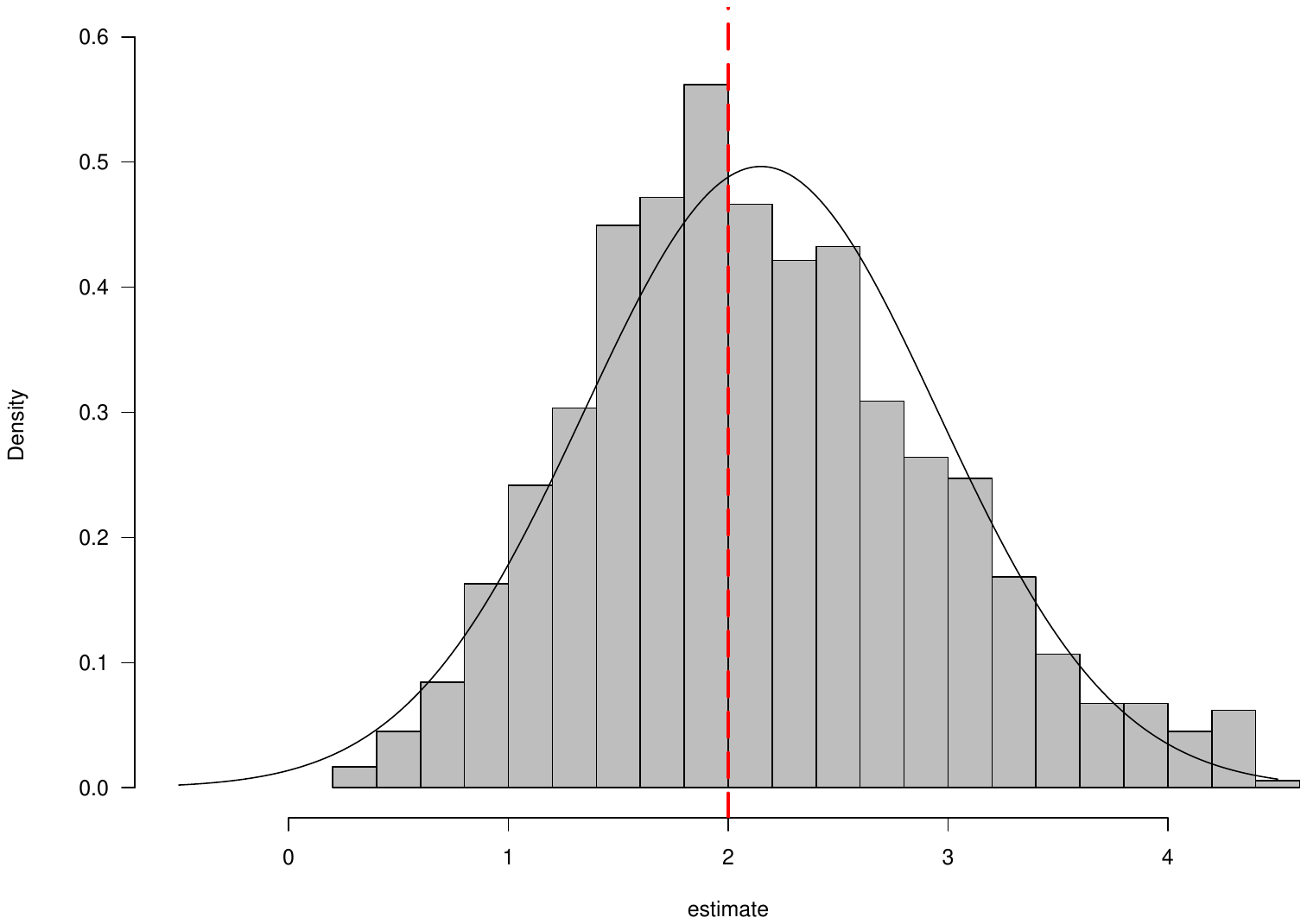}}
    \newline
  \centering
  \subfigure[\small $\hat{\beta}^\ast_{\text{semi}}$: incorrectly specified $U$, $n = 500$]{%
     \label{app fig: binary U binary Y incorrect U n = 500}%
    \includegraphics[height = 5 cm, width = 0.5\linewidth]{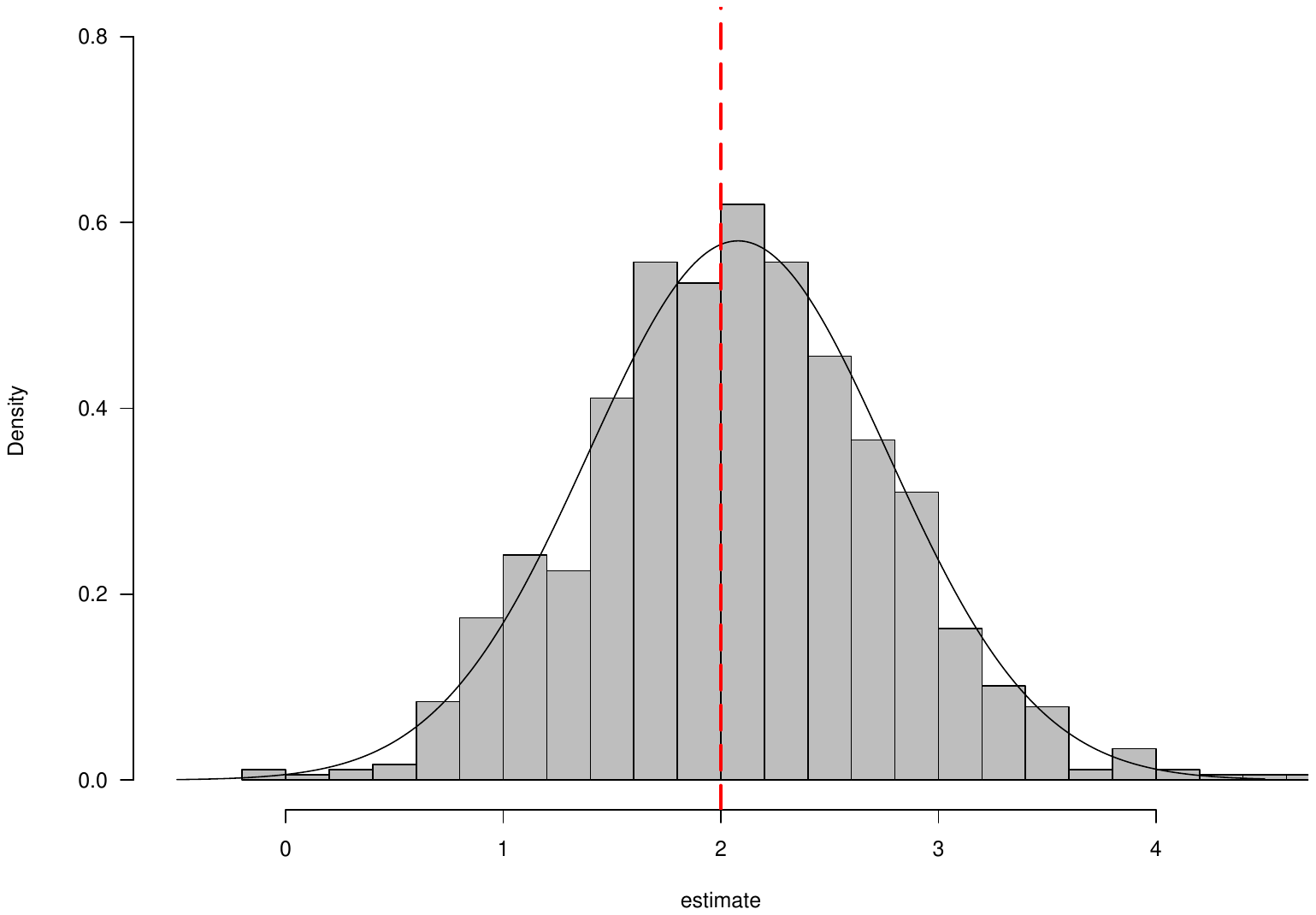}}%
  \subfigure[\small $\hat{\beta}_{\text{semi}}$: correctly specified $U$, $n = 500$]{%
    \label{app fig: binary U binary Y correct U n = 500}%
    \includegraphics[height = 5 cm, width = 0.5\linewidth]{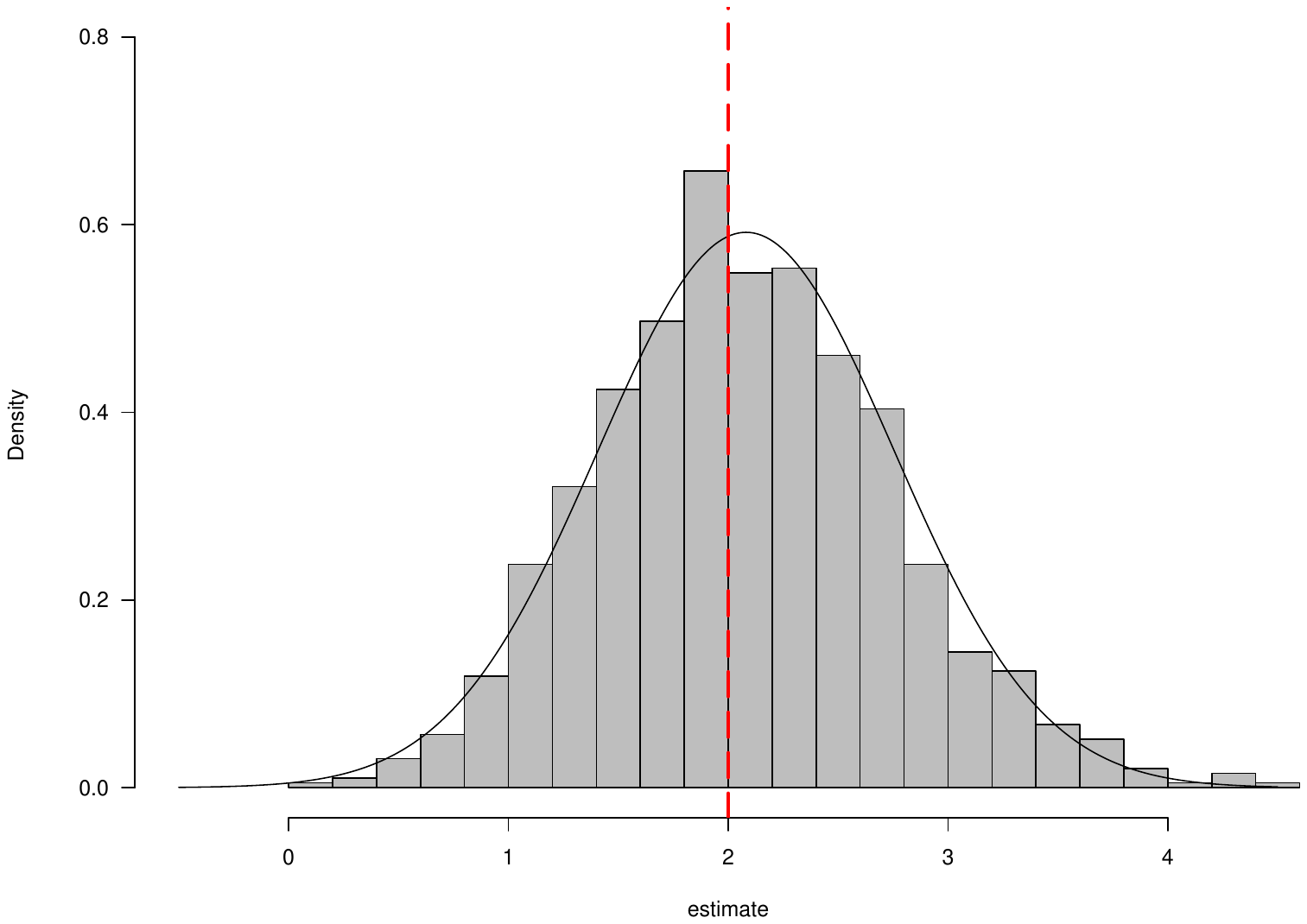}} 
  \caption{\small Semiparametric estimators when both U and Y are binary. Top two panels: $n = 300$. Bottom two panels: $n = 500$. True $\beta$ value is represented by a red vertical line in all panels.}
  \label{appendix fig: binary U binary Y n = 300, 500}
\end{figure}

\clearpage
\subsection*{E.2: Binary U and Continuous Y: Plots of Monte Carlo Distributions}
\begin{figure}[h]%
  \centering
  \subfigure[\small $\hat{\beta}_{\text{semi}}$: correctly specified $U$]{%
     \label{fig: binary U cont Y incorrect U}%
    \includegraphics[height = 10 cm, width = 0.5\linewidth]{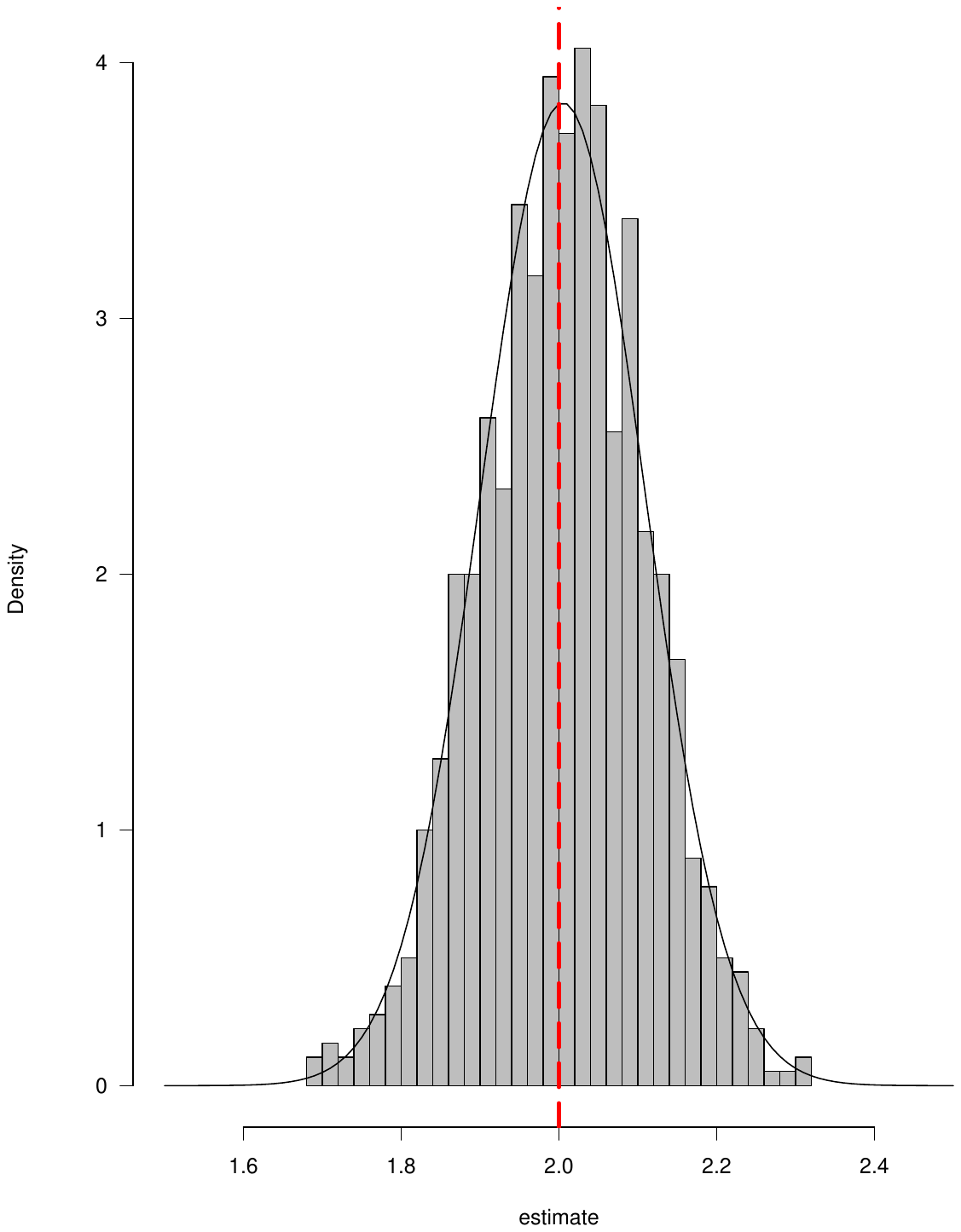}}%
  \subfigure[\small $\hat{\beta}^\ast_{\text{semi}}$: incorrectly specified $U$]{%
    \label{fig: binary U cont Y correct U}%
    \includegraphics[height = 10 cm, width = 0.5\linewidth]{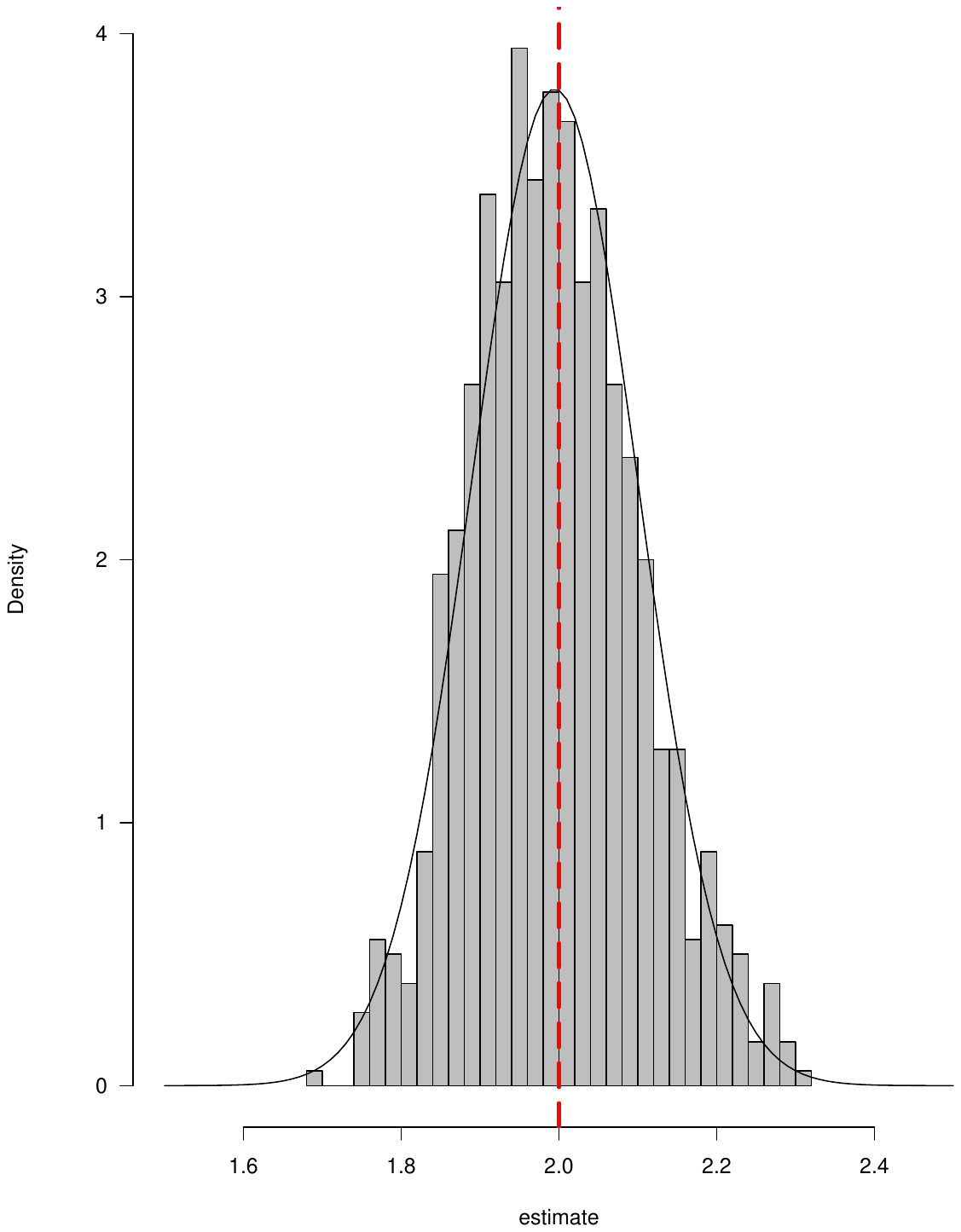}}
  \caption{\small Two semiparametric estimators when $U$ is binary and $Y$ is continuous. True $\beta$ value is represented by a red vertical line in both panels.}
  \label{fig: two semi estimators cont y}
\end{figure}

We also considered a continuous $Y$ and a binary $U$. We specified the following DGP: 
\begin{equation}
    \begin{split}
        &X_1 \sim \text{Uniform}(0, 1); ~ X_2 \sim \text{Uniform}(0, 1) \\
        &U \sim \text{Bernoulli}(0.2) \\
        &\text{logit}(Z \mid X_1, X_2, U) = 3X_1 - 3X_2 + \lambda U \\
        &Y = X_1 + X_2 + 2 Z + c_\delta U + \epsilon, \quad \epsilon \sim N(0, 1),
    \end{split}
\end{equation}
where $\lambda = c_\delta = 4$ and $n = 500$  When $Y$ is continuous, we approximate the kernel function in Theorem \ref{thm: fredholm equivalence} using Hermite quadrature. Figure \ref{fig: two semi estimators cont y} plots the Monte Carlo distributions of $\hat{\beta}_{\text{semi}}$ and $\hat{\beta}^\ast_{\text{semi}}$, two semiparametric estimators with a correctly and an incorrectly specified $U$, respectively.

\clearpage
\bibliographystyle{rss}
\bibliography{semi}

\begin{thebibliography}{73}
\expandafter\ifx\csname natexlab\endcsname\relax\def\natexlab#1{#1}\fi
\expandafter\ifx\csname url\endcsname\relax
  \def\url#1{\texttt{#1}}\fi
\expandafter\ifx\csname urlprefix\endcsname\relax\def\urlprefix{URL: }\fi

\bibitem[{Allen et~al.(2005)Allen, Satten and Tsiatis}]{allen2005locally}
Allen, A.~S., Satten, G.~A. and Tsiatis, A.~A. (2005) Locally-efficient robust
  estimation of haplotype-disease association in family-based studies.
\newblock \textit{Biometrika}, \textbf{92}, 559--571.

\bibitem[{Altonji et~al.(2005)Altonji, Elder and Taber}]{altonji2005selection}
Altonji, J.~G., Elder, T.~E. and Taber, C.~R. (2005) Selection on observed and
  unobserved variables: Assessing the effectiveness of catholic schools.
\newblock \textit{Journal of political economy}, \textbf{113}, 151--184.

\bibitem[{Annan et~al.(2006)Annan, Blattman and Horton}]{Annan2006}
Annan, J., Blattman, C. and Horton, R. (2006) The state of youth and youth
  protection in northern uganda.
\newblock \textit{Uganda: UNICEF}, \textbf{23}.

\bibitem[{Baker et~al.(1964)Baker, Fox, Mayers and Wright}]{baker1964numerical}
Baker, C.~T., Fox, L., Mayers, D. and Wright, K. (1964) Numerical solution of
  fredholm integral equations of first kind.
\newblock \textit{The Computer Journal}, \textbf{7}, 141--148.

\bibitem[{Bang and Robins(2005)}]{Bang2005}
Bang, H. and Robins, J.~M. (2005) Doubly robust estimation in missing data and
  causal inference models.
\newblock \textit{Biometrics}, \textbf{61}, 962--973.

\bibitem[{Barnow et~al.(1980)Barnow, G.G.Cain and A.S.Goldberg}]{Barnow1980}
Barnow, B., G.G.Cain and A.S.Goldberg (1980) Issues in the analysis of
  selectivity bias.
\newblock In \textit{Evaluation Studies, Volume 5} (eds. E.Stromsdorfer and
  G.Farkas). San Francisco, CA: Sage.

\bibitem[{Bickel et~al.(1993)Bickel, Klaassen, Wellner and
  Ritov}]{Bickel1993efficient}
Bickel, P.~J., Klaassen, C.~A., Wellner, J.~A. and Ritov, Y. (1993)
  \textit{Efficient and adaptive estimation for semiparametric models}, vol.~4.
\newblock Johns Hopkins University Press Baltimore.

\bibitem[{Blattman(2009)}]{Blattman2009}
Blattman, C. (2009) {From violence to voting: war and political participation
  in Uganda}.
\newblock \textit{American Political Science Review}, \textbf{103}, 231–247.

\bibitem[{Blattman and Annan(2010)}]{Blattman2010}
Blattman, C. and Annan, J. (2010) The consequences of child soldiering.
\newblock \textit{The Review of Economics and Statistics}, \textbf{92},
  882--898.

\bibitem[{Carnegie et~al.(2016)Carnegie, Harada and
  Hill}]{carnegie2016assessing}
Carnegie, N.~B., Harada, M. and Hill, J.~L. (2016) Assessing sensitivity to
  unmeasured confounding using a simulated potential confounder.
\newblock \textit{Journal of Research on Educational Effectiveness},
  \textbf{9}, 395--420.

\bibitem[{Carrasco et~al.(2007)Carrasco, Florens and
  Renault}]{carrasco2007linear}
Carrasco, M., Florens, J.-P. and Renault, E. (2007) Linear inverse problems in
  structural econometrics estimation based on spectral decomposition and
  regularization.
\newblock \textit{Handbook of Econometrics}, \textbf{6}, 5633--5751.

\bibitem[{Cinelli and Hazlett(2020)}]{Cinelli2020}
Cinelli, C. and Hazlett, C. (2020) Making sense of sensitivity: extending
  omitted variable bias.
\newblock \textit{Journal of the Royal Statistical Society: Series B
  (Statistical Methodology)}, \textbf{82}, 39--67.

\bibitem[{Collier(2007)}]{Collier2008}
Collier, P. (2007) \textit{The Bottom Billion}.
\newblock Oxford: Oxford University Press.

\bibitem[{Copas and Li(1997)}]{Copas1997}
Copas, J.~B. and Li, H.~G. (1997) Inference for non-random samples.
\newblock \textit{Journal of the Royal Statistical Society. Series B
  (Methodological)}, \textbf{59}, 55--95.

\bibitem[{Cornfield et~al.(1959)Cornfield, Haenszel, Hammond, Lilienfeld,
  Shimkin and Wynder}]{Cornfield1959}
Cornfield, J., Haenszel, W., Hammond, E., Lilienfeld, A., Shimkin, M. and
  Wynder, E. (1959) {Smoking and lung cancer}.
\newblock \textit{Journal of the National Cancer Institute}, \textbf{22},
  173--203.

\bibitem[{Ding and VanderWeele(2016)}]{ding2016sensitivity}
Ding, P. and VanderWeele, T.~J. (2016) Sensitivity analysis without
  assumptions.
\newblock \textit{Epidemiology (Cambridge, Mass.)}, \textbf{27}, 368.

\bibitem[{DiPrete and Gangl(2004)}]{DiPrete2004}
DiPrete, T.~A. and Gangl, M. (2004) {Assessing bias in the estimation of causal
  effects: Rosenbaum bounds on matching estimators and instrumental variables
  estimation with imperfect instruments}.
\newblock \textit{Sociological methodology}, \textbf{34}, 271--310.

\bibitem[{Dorie et~al.(2016)Dorie, Harada, Carnegie and Hill}]{Dorie2016}
Dorie, V., Harada, M., Carnegie, N.~B. and Hill, J. (2016) A flexible,
  interpretable framework for assessing sensitivity to unmeasured confounding.
\newblock \textit{Statistics in Medicine}, \textbf{35}, 3453--3470.

\bibitem[{Fisher(1958)}]{fisher1958cancer}
Fisher, R.~A. (1958) Cancer and smoking.
\newblock \textit{Nature}, \textbf{182}, 596--596.

\bibitem[{Foutz(1977)}]{Foutz1977}
Foutz, R.~V. (1977) On the unique consistent solution to the likelihood
  equations.
\newblock \textit{Journal of the American Statistical Association},
  \textbf{72}, 147--148.

\bibitem[{Franks et~al.(2019)Franks, D’Amour and Feller}]{franks2019flexible}
Franks, A., D’Amour, A. and Feller, A. (2019) Flexible sensitivity analysis
  for observational studies without observable implications.
\newblock \textit{Journal of the American Statistical Association}.

\bibitem[{Garcia and Ma(2016)}]{garcia2016optimal}
Garcia, T.~P. and Ma, Y. (2016) Optimal estimator for logistic model with
  distribution-free random intercept.
\newblock \textit{Scandinavian Journal of Statistics}, \textbf{43}, 156--171.

\bibitem[{Gastwirth et~al.(1998)Gastwirth, Krieger and
  Rosenbaum}]{Gastwirth1998}
Gastwirth, J., Krieger, A.~M. and Rosenbaum, P.~R. (1998) {Dual and
  simultaneous sensitivity analysis for matched pairs}.
\newblock \textit{Biometrika}, \textbf{85}, 907--920.

\bibitem[{Greenland and Robins(1986)}]{Greenland1986}
Greenland, S. and Robins, J.~M. (1986) {Identifiability, exchangeability, and
  epidemiological confounding}.
\newblock \textit{International Journal of Epidemiology}, \textbf{15},
  413--419.

\bibitem[{Griffin et~al.(2013)Griffin, Eibner, Bird, Jewell, Margolis, Shih,
  Slaughter, Whitsel, Allison and Escarce}]{Griffin2013}
Griffin, B.~A., Eibner, C., Bird, C.~E., Jewell, A., Margolis, K., Shih, R.,
  Slaughter, M.~E., Whitsel, E.~A., Allison, M. and Escarce, J.~J. (2013) The
  relationship between urban sprawl and coronary heart disease in women.
\newblock \textit{Health \& Place}, \textbf{20}, 51 -- 61.

\bibitem[{Hill(2011)}]{Hill2011}
Hill, J.~L. (2011) Bayesian nonparametric modeling for causal inference.
\newblock \textit{Journal of Computational and Graphical Statistics},
  \textbf{20}, 217--240.

\bibitem[{Ho et~al.(2007)Ho, Imai, King and Stuart}]{ho2007matching}
Ho, D.~E., Imai, K., King, G. and Stuart, E.~A. (2007) Matching as
  nonparametric preprocessing for reducing model dependence in parametric
  causal inference.
\newblock \textit{{Political Analysis}}, \textbf{15}, 199--236.

\bibitem[{Hsu and Small(2013)}]{Hsu2013}
Hsu, J.~Y. and Small, D.~S. (2013) {Calibrating sensitivity analyses to
  observed covariates in observational studies}.
\newblock \textit{Biometrics}, \textbf{69}, 803--811.

\bibitem[{Ichino et~al.(2008)Ichino, Mealli and Nannicini}]{Ichino2008}
Ichino, A., Mealli, F. and Nannicini, T. (2008) From temporary help jobs to
  permanent employment: what can we learn from matching estimators and their
  sensitivity?
\newblock \textit{Journal of Applied Econometrics}, \textbf{23}, 305--327.

\bibitem[{Imbens(2003)}]{Imbens2003}
Imbens, G.~W. (2003) {Sensitivity to exogeneity assumptions in program
  evaluation}.
\newblock \textit{American Economic Review}, \textbf{93}, 126--132.

\bibitem[{Imbens(2004)}]{Imbens2004}
--- (2004) {Nonparametric estimation of average treatment effects under
  exogeneity: A review}.
\newblock \textit{Review of Economics and Statistics}, \textbf{86}, 4--29.

\bibitem[{Kress et~al.(1989)Kress, Maz'ya and Kozlov}]{kress1989linear}
Kress, R., Maz'ya, V. and Kozlov, V. (1989) \textit{Linear integral equations},
  vol.~82.
\newblock Springer.

\bibitem[{McCandless et~al.(2007)McCandless, Gustafson and
  Levy}]{McCandless2007}
McCandless, L.~C., Gustafson, P. and Levy, A. (2007) {Bayesian sensitivity
  analysis for unmeasured confounding in observational studies}.
\newblock \textit{Statistics in Medicine}, \textbf{26}, 2331–2347.

\bibitem[{Newey(1990)}]{Newey1990}
Newey, W.~K. (1990) Semiparametric efficiency bounds.
\newblock \textit{Journal of Applied Econometrics}, \textbf{5}, 99--135.

\bibitem[{Neyman(1923)}]{Neyman1923}
Neyman, J. (1923) {On the application of probability theory to agricultural
  experiments}.
\newblock \textit{Reprint in Statistical Science}, \textbf{5}, 465--480.

\bibitem[{Phillips(1962)}]{phillips1962technique}
Phillips, D.~L. (1962) A technique for the numerical solution of certain
  integral equations of the first kind.
\newblock \textit{Journal of the ACM (JACM)}, \textbf{9}, 84--97.

\bibitem[{Robins(1986)}]{Robins1986}
Robins, J.~M. (1986) A new approach to causal inference in mortality studies
  with a sustained exposure period—application to control of the healthy
  worker survivor effect.
\newblock \textit{Mathematical Modelling}, \textbf{7}, 1393 -- 1512.

\bibitem[{Robins(1992)}]{Robins1992}
--- (1992) { Estimation of the time-dependent accelerated failure time model in
  the presence of confounding factors}.
\newblock \textit{Biometrika}, \textbf{79}, 321--334.

\bibitem[{Robins(2000)}]{Robins2000b}
--- (2000) Robust estimation in sequentially ignorable missing data and causal
  inference models.
\newblock \textit{ASA Proceedings of the Section on Bayesian Statistical
  Science}, \textbf{1999}.

\bibitem[{Robins et~al.(2000)Robins, Ángel Hernán and Brumback}]{Robins2000c}
Robins, J.~M., Ángel Hernán, M. and Brumback, B. (2000) Marginal structural
  models and causal inference in epidemiology.
\newblock \textit{Epidemiology}, \textbf{11}, 550--560.

\bibitem[{Robins et~al.(1994)Robins, Rotnitzky and Zhao}]{Robins1994}
Robins, J.~M., Rotnitzky, A. and Zhao, L.~P. (1994) Estimation of regression
  coefficients when some regressors are not always observed.
\newblock \textit{Journal of the American Statistical Association},
  \textbf{89}, 846--866.

\bibitem[{Rosenbaum(1987{\natexlab{a}})}]{Rosenbaum1987b}
Rosenbaum, P.~R. (1987{\natexlab{a}}) Model-based direct adjustment.
\newblock \textit{Journal of the American Statistical Association},
  \textbf{82}, 387--394.

\bibitem[{Rosenbaum(1987{\natexlab{b}})}]{Rosenbaum1987}
--- (1987{\natexlab{b}}) {Sensitivity analysis for certain permutation
  inferences in matched observational studies}.
\newblock \textit{Biometrika}, \textbf{74}, 13--26.

\bibitem[{Rosenbaum(1987{\natexlab{c}})}]{rosenbaum1987sensitivity}
--- (1987{\natexlab{c}}) Sensitivity analysis for certain permutation
  inferences in matched observational studies.
\newblock \textit{Biometrika}, \textbf{74}, 13--26.

\bibitem[{Rosenbaum(1989)}]{rosenbaum1989sensitivity}
--- (1989) Sensitivity analysis for matched observational studies with many
  ordered treatments.
\newblock \textit{Scandinavian Journal of Statistics}, 227--236.

\bibitem[{Rosenbaum(2002)}]{Rosenbaum2002a}
--- (2002) \textit{Observational Studies}.
\newblock Springer.

\bibitem[{Rosenbaum(2010)}]{Rosenbaum2010}
--- (2010) \textit{Design of Observational Studies}.
\newblock Springer, New York.

\bibitem[{Rosenbaum and Rubin(1983{\natexlab{a}})}]{Rosenbaum1983}
Rosenbaum, P.~R. and Rubin, D.~B. (1983{\natexlab{a}}) {Assessing sensitivity
  to an unobserved binary covariate in an observational study with binary
  outcome}.
\newblock \textit{Journal of Royal Statistical Society, Series B}, \textbf{45},
  212--218.

\bibitem[{Rosenbaum and Rubin(1983{\natexlab{b}})}]{rosenbaum1983central}
--- (1983{\natexlab{b}}) The central role of the propensity score in
  observational studies for causal effects.
\newblock \textit{Biometrika}, \textbf{70}, 41--55.

\bibitem[{Rosenbaum and Rubin(1984)}]{Rosenbaum1984}
--- (1984) Reducing bias in observational studies using subclassification on
  the propensity score.
\newblock \textit{Journal of the American Statistical Association},
  \textbf{79}, 516--524.

\bibitem[{Rosenbaum and Silber(2009)}]{Rosenbaum2009}
Rosenbaum, P.~R. and Silber, J.~H. (2009) {Amplification of sensitivity
  analysis in matched observational studies}.
\newblock \textit{Journal of the American Statistical Association},
  \textbf{104}, 1398--1405.

\bibitem[{Rosenbaum and Small(2017)}]{rosenbaum2017adaptive}
Rosenbaum, P.~R. and Small, D.~S. (2017) An adaptive mantel--haenszel test for
  sensitivity analysis in observational studies.
\newblock \textit{Biometrics}, \textbf{73}, 422--430.

\bibitem[{Rubin(1987)}]{Rubin1987}
Rubin, D. (1987) \textit{{Multiple Imputation for Nonresponse in Surveys}}.
\newblock New York: Wiley.

\bibitem[{Rubin(1974)}]{Rubin1974}
Rubin, D.~B. (1974) {Estimating causal effects of treatments in randomized and
  nonrandomized studies}.
\newblock \textit{Journal of Educational Psychology}, \textbf{66}, 688--701.

\bibitem[{Rubin(1979)}]{Rubin1979}
--- (1979) {Using multivariate matched sampling and regression adjustment to
  control bias in observational studies}.
\newblock \textit{Journal of the American Statistical Association},
  \textbf{74}, 318--328.

\bibitem[{Rubin(1980)}]{Rubin1980}
--- (1980) {Randomization analysis of experimental data: the Fisher
  randomization test comment}.
\newblock \textit{Journal of the American Statistical Association},
  \textbf{75}, 591--593.

\bibitem[{Scharfstein et~al.(1999)Scharfstein, Rotnitzky and
  Robins}]{Scharfstein1999}
Scharfstein, D.~O., Rotnitzky, A. and Robins, J.~M. (1999) Adjusting for
  nonignorable drop-out using semiparametric nonresponse models.
\newblock \textit{Journal of the American Statistical Association},
  \textbf{94}, 1096--1120.

\bibitem[{Soetaert and Herman(2009)}]{R_rootSolve}
Soetaert, K. and Herman, P.~M. (2009) \textit{A Practical Guide to Ecological
  Modelling. Using R as a Simulation Platform}.
\newblock Springer.
\newblock ISBN 978-1-4020-8623-6.

\bibitem[{Spear(2016)}]{spear2016disarmament}
Spear, J. (2016) {Disarmament, demobilization, reinsertion and reintegration in
  Africa}.
\newblock In \textit{Ending Africa's wars}, 73--90. Routledge.

\bibitem[{Stuart(2010)}]{Stuart2010}
Stuart, E.~A. (2010) {Matching methods for causal inference: a review and a
  look forward}.
\newblock \textit{Statistical Science}, \textbf{25}, 1--21.

\bibitem[{Tikhonov(1963)}]{tikhonov1963solution}
Tikhonov, A.~N. (1963) On the solution of ill-posed problems and the method of
  regularization.
\newblock In \textit{Doklady Akademii Nauk}, vol. 151, 501--504. Russian
  Academy of Sciences.

\bibitem[{Tsiatis(2006)}]{Tsiatis2006}
Tsiatis, A. (2006) \textit{{Semiparametric Theory and Missing Data}}.
\newblock New York: Springer.

\bibitem[{Tsiatis and Ma(2004)}]{Tsiatis2004}
Tsiatis, A.~A. and Ma, Y. (2004) {Locally efficient semiparametric estimators
  for functional measurement error models}.
\newblock \textit{Biometrika}, \textbf{91}, 835--848.

\bibitem[{Van~der Vaart(2000)}]{van2000asymptotic}
Van~der Vaart, A.~W. (2000) \textit{Asymptotic statistics}, vol.~3.
\newblock Cambridge university press.

\bibitem[{van~der Vaart and Wellner(1996)}]{vaart1996weak}
van~der Vaart, A.~W. and Wellner, J.~A. (1996) \textit{Weak convergence and
  empirical processes: with applications to statistics}.
\newblock Springer.

\bibitem[{{van Buuren} and Groothuis-Oudshoorn(2011)}]{R_mice}
{van Buuren}, S. and Groothuis-Oudshoorn, K. (2011) {mice}: Multivariate
  imputation by chained equations in r.
\newblock \textit{Journal of Statistical Software}, \textbf{45}, 1--67.

\bibitem[{VanderWeele and Arah(2011)}]{VanderWeele2011}
VanderWeele, T.~J. and Arah, O.~A. (2011) {Bias formulas for sensitivity
  analysis of unmeasured confounding for general outcomes, treatments, and
  confounders}.
\newblock \textit{Epidemiology}, \textbf{22}, 42--52.

\bibitem[{Vansteelandt and Joffe(2014)}]{Vansteelandt2014}
Vansteelandt, S. and Joffe, M. (2014) Structural nested models and
  g-estimation: The partially realized promise.
\newblock \textit{Statistical Science}, \textbf{29}, 707--731.

\bibitem[{Wasserman(1999)}]{Wasserman1999}
Wasserman, L. (1999) Estimation of the causal effect of a time-varying exposure
  on the marginal mean of a repeated binary outcome: Comment.
\newblock \textit{Journal of the American Statistical Association},
  \textbf{94}, 704--706.

\bibitem[{Wooldridge(2008)}]{Wooldridge2008}
Wooldridge, J. (2008) \textit{{Introductory Econometrics: A Modern Approach
  (with Economic Applications, Data Sets, Student Solutions Manual Printed
  Access Card)}}.
\newblock South-Western College Pub.

\bibitem[{Zhang and Small(2020)}]{zhang2018calibrated}
Zhang, B. and Small, D.~S. (2020) A calibrated sensitivity analysis for matched
  observational studies with application to the effect of second-hand smoke
  exposure on blood lead levels in children.
\newblock \textit{Journal of the Royal Statistical Society: Series C (Applied
  Statistics)}, \textbf{69}, 1285--1305.

\bibitem[{Zhao(2019)}]{zhao2019sensitivity}
Zhao, Q. (2019) On sensitivity value of pair-matched observational studies.
\newblock \textit{Journal of the American Statistical Association},
  \textbf{114}, 713--722.

\bibitem[{Zhao et~al.(2019)Zhao, Small and Bhattacharya}]{zhao_sens_ipw2019}
Zhao, Q., Small, D.~S. and Bhattacharya, B.~B. (2019) Sensitivity analysis for
  inverse probability weighting estimators via the percentile bootstrap.
\newblock \textit{Journal of the Royal Statistical Society: Series B
  (Statistical Methodology)}, \textbf{0}.

\end{thebibliography}
\end{document}